\documentclass{article} 
\usepackage{iclr2027_conference}
\usepackage{times}

\usepackage{amsmath,amsfonts,bm}

\def\eqref#1{equation~\ref{#1}}

\def\1{\bm{1}}

\DeclareMathAlphabet{\mathsfit}{\encodingdefault}{\sfdefault}{m}{sl}
\SetMathAlphabet{\mathsfit}{bold}{\encodingdefault}{\sfdefault}{bx}{n}

\usepackage[hidelinks]{hyperref}
\usepackage{url}
\usepackage{booktabs}
\usepackage{graphicx}
\usepackage{wrapfig}
\usepackage{xcolor}
\usepackage{subcaption}
\usepackage{MnSymbol}
\usepackage{aas_macros}

\usepackage[leftcaption]{sidecap}


\usepackage{multirow}

\title{The Platonic Universe: \\Do Foundation Models See the Same Sky?}

\author{%
\begin{tabular}{l}
\\[8pt]
UniverseTBD\footnotemark\\[8pt] 
Trinidad Borrell$^{1,2}$, Steven Dillmann$^3$, Kshitij Duraphe$^4$, Furkan Eris$^5$
\\[4pt] Kartheik Iyer$^6$, Ashod Khederlarian$^7$, Aman Kumar$^8$, Giovanni Marraffini$^{9,10}$ \\[4pt] Michael J. Smith$^{11,12}$, Shashwat Sourav$^{13,14,15}$, Rocco Di Tella$^{11,12}$, John F. Wu$^{16,17}$ 
 \\[8pt]
\normalfont $^1$Paris Brain Institute, $^2$Sorbonne Universit\'e, $^3$Stanford University, $^4$Independent Researcher\\
\normalfont $^5$Prolific, $^6$University of Hawai'i ,$^7$University of Pittsburgh, $^8$IUCAA, $^9$INRIA, $^{10}$Sigma Nova\\ 
\normalfont $^{11}$AstroAI, $^{12}$Center for Astrophysics $\vert$ Harvard \& Smithsonian\\
\normalfont $^{13}$Washington University in St.\ Louis,
\normalfont $^{14}$Oak Ridge National Lab\\
\normalfont $^{15}$Lawrence Berkeley National Lab, $^{16}$Space Telescope Science Institute\\
\normalfont$^{16}$Johns Hopkins University
\normalfont
\end{tabular}
}

\iclrfinalcopy

\begin{document}

\maketitle

\begin{abstract}
We investigate when foundation models converge towards shared representations, and how this convergence depends on model capacity, training regime, and model architecture.
We take a `science-for-AI' approach, using astronomy as an experimental instrument to test the Platonic Representation Hypothesis (PRH) and its Aristotelian refinement (ARH) against an external physical reference.
The historical success of astrophysics is evidence that a compact, modality-invariant description of galaxy observables exists, and so representation convergence toward reality should be measurable against the physical parameters astronomers already use.
Given this framework, we evaluate eleven foundation model families (spanning supervised classification, self-distillation, joint-embedding prediction, masked autoencoding, vision-language pre-training, and astronomy-specific architectures from $\mathcal{O}$(10M)${\to}\mathcal{O}$(10B)  parameters)  on crossmatched JWST, HSC, and Legacy Survey imagery, and DESI spectroscopy.
All models are evaluated frozen, with no astronomy-specific fine-tuning.
We probe redshift, stellar mass, and specific star formation rate via linear probes, and local (MKNN) and global (CKA) embedding geometry within families, between modalities, and across architectures.
We find that physics performance scales predictably with capacity; probe directions align consistently with expected astrophysical correlations and selection effects; and local (but not global) embedding alignment tracks physics performance, including between DESI spectra and HSC imagery---modalities that share essentially no low-level statistics.
Our results support the ARH over the strict PRH, demonstrate astronomy's value as an experimental framework for neural representation learning, and suggest that astro-foundation models can build on general-purpose pre-trained architectures, capitalizing on the broader open machine learning community's already-spent computational investment.
\end{abstract}

\begingroup
\renewcommand{\thefootnote}{\fnsymbol{footnote}}
\footnotetext[1]{Alphabetical ordering, author contributions are stated after the acknowledgements.}
\endgroup

\section{Astronomy and the Platonic Representation Hypothesis}

Three historical waves of increasingly automated connectionism have swept the shores of astronomy. 
The late 1980s brought MLPs tuned for astronomical applications on manually selected inputs \citep[e.g.][]{ref_adorf1988,ref_angel1990,ref_odewahn1992}.
With the advent of CNNs, RNNs, and deep learning, these MLP models gave way to raw data ingestion \citep[e.g.][]{ref_dieleman2015,ref_charnock2018,ref_wu2020,ref_smith2021p2p,2026arXiv260210207K}.
And the third wave of unsupervised and self-supervised learning largely removed the need for task-specific human-generated labelling, with connectionist methods inferring astronomical knowledge directly from the raw data \citep[e.g.][]{ref_sarmiento2021,ref_smith2022}.
A fourth wave has recently been seeded by the discovery of predictable neural scaling laws \citep{ref_kaplan2020,ref_hoffmann2022}---the application of foundation models to astronomical observations, publications, and survey data \citep{ref_smith2023}.
This fourth wave has brought with it diverse approaches in the search for a viable path towards a single, canonical, astro-foundation model: from contrastive methods \citep[e.g.][]{ref_slijepcevic2024,ref_parker2024,ref_mishra2024,ref_zhao2025}, to generative  architectures  \citep[e.g.][]{ref_leung2023,ref_koblischke2024,ref_ore2024}, to autoregressive modelling \citep[e.g.][]{ref_smith2024,ref_pan2024,ref_euclid2025,ref_zuo2025,ref_heneka2025,ref_moriwaki2025}, to finetuning of large language models on astronomical text \citep[e.g.][]{ref_nguyen2023,ref_perkowski2024,ref_dehaan2024,ref_zaman2025}.
The sheer diversity of these approaches raises a natural question: does the choice of architecture, training regime, and data modality matter, or do sufficiently performant models converge to equivalent representations?

The Platonic Representation Hypothesis (PRH; \citet{ref_huh2024}) offers one answer: neural networks trained on different modalities are converging toward a shared global statistical model of reality in their representation spaces.
The PRH attributes this convergence to three mechanisms---\textit{task generality} (models trained on more diverse tasks require representations that capture more information about underlying reality), \textit{model capacity} (larger models are more likely to find optimal representations), and \textit{simplicity bias} (neural networks naturally favour simpler solutions that generalise better)---and draws inspiration from Plato's Allegory of the Cave \citep{ref_plato-375}, in which the cave-dwellers mistake shadows on a wall for reality itself.
In this analogy, our training data are the shadowy projections of an underlying reality, and our models learn to recover representations (or `Forms') of the reality that generates them. 
Under the PRH, larger models trained on more diverse tasks should converge toward a Platonic Ideal: a perfect, lossless Form of underlying reality.
A recent refinement, the Aristotelian Representation Hypothesis (ARH; \citet{ref_groger2026}), tempers this strong claim. 
Rather than agreeing on a shared global coordinate system, the ARH proposes that converging models agree on local neighborhood structure  rather than global embedding geometry. 
This distinction has direct empirical consequences in that \emph{local} rank statistics are predicted to track convergence, while \emph{global} similarity measures are not.

We use astronomy as our experimental instrument.
Astronomical observations provide a natural testbed as different projections of the same underlying cosmic reality.
The historical success of astrophysics is itself evidence that a compact, modality-invariant description of galaxy observables exists, in that distinct instruments observing the same galaxies can be reconciled into a single physical account.
These observations are linked through shared physical processes;
a galaxy's morphology, kinematics, chemical composition, and star formation history all emerge from the same stellar populations, gas and dust dynamics, and underlying matter distributions \citep{ref_conroy2013}.
This provides an external physical reference against which we can measure whether foundation models converge towards representations of the underlying physics.
Furthermore, the scale and diversity of modern surveys provide the data volume necessary to test convergence across multiple model architectures and training objectives, and recent work \citep{ref_mmu2024,ref_caplar2025} has standardized crossmatching across astronomical modes\footnote{
    We define an astronomical `mode' or `modality' as the information captured by a specific instrument so that (e.g.) JWST and HSC imaging are separate modes.
}.

Our contributions follow.
\textbf{We establish the use of astronomy as a instrument for the study of neural representations} by testing both the PRH and the ARH across a basket of eleven foundation model families spanning supervised classification, self-distillation, joint-embedding prediction, masked autoencoding, vision–language pre-training, and astronomy-specific architectures from $\mathcal{O}$(10M)${\to}\mathcal{O}$(10B) parameters, evaluated on crossmatched JWST, HSC, and Legacy Survey imaging, and DESI spectroscopy. 
We organise our analysis around answering four questions:
\textbf{(1) Do larger models encode more galaxy physics?} We train linear probes on frozen embeddings to predict redshift, stellar mass, and specific star formation rate given galaxy images, and find that probe performance scales with model capacity (\S\ref{subsec_probes} \P1). 
\textbf{(2) Do different architectures converge to the same physical representation?} We find that physics performance is correlated between modalities, and the relative orientations of probe directions agree across models and with expected astrophysical relationships (\S\ref{subsec_probes} \P2-3).
\textbf{(3) Does geometric alignment scale with performance---locally, globally, or both?}
We find that MKNN (local alignment) but not CKA (global alignment) tracks probe performance across model families, between astronomical modalities, and across architectures (\S\ref{subsec_mknncka}).
\textbf{(4) What can we learn from our anomalous results?}
We find that DINOv3 and PaliGemma do not follow the trend established by other models:
larger DINOv3 students align more closely with their fixed teacher but less with an average model, while scaling PaliGemma’s language trunk around a fixed vision encoder does not improve physics performance (\S\ref{subsec_mknncka}; App.~\ref{app_anomaly}), suggesting that total model capacity alone is insufficient to predict representational convergence.

\section{Datasets, models, and metrics} \label{sec_methods}


\textbf{Datasets used.} We test across four crossmatched astronomical datasets that capture different projections of galaxy Forms.
We use the MMU to crossmatch between data modes at a 1" radius \citep{ref_mmu2024}.
\textbf{Hyper Suprime-Cam (HSC)} \citep{ref_miyazaki2018} provides our reference baseline. HSC is an 8.2m ground-based optical survey with ${\sim}$0.7" seeing and 5-band (\emph{grizy}) coverage. We use HSC as it provides our largest overlap sample across all other modalities, making it a natural anchor for pairwise comparison.
\textbf{DESI Legacy Imaging Survey} \citep{ref_dey2019} is a complementary ground-based optical survey at ${\sim}1"$ seeing, its inclusion allows us to test representational alignment across different ground-based imaging survey strategies. Our crossmatched Legacy-HSC sample contains 102k galaxies.
\textbf{JWST NIRCam imaging} \citep{ref_jwst2023} provides our most extreme imaging comparison: space-based near-infrared and infrared observations that can reveal dust emission and dust-obscured stellar populations invisible to our optical surveys.
The JWST-HSC crossmatched sample contains 1.67k galaxies.
\textbf{DESI spectra} \citep{ref_desi2024} provide our crossmodal non-imaging test. 
Spectroscopy probes integrated galaxy properties (e.g. stellar properties, kinematics) that are complementary to the spatial information captured by imaging. 
The DESI-HSC crossmatched sample contains 18.6k galaxies. 

To test the convergence of representations with astronomical parameters, we use HSC and JWST images from \textbf{COSMOS-Web} \citep{2023ApJ...954...31C}. The COSMOS field is one of the most observed areas of the night sky; due to imaging from many different telescopes across wide wavelength ranges, galaxies in COSMOS have very accurate measurements of their physical parameters \citep{2025A&A...704A.339S}. 
To ensure that our physical labels are high quality, we select 45\,000 galaxies with both HSC $i$ and JWST F150W band flux signal-to-noise ratios greater than 25, and use their redshifts (many-band photometric redshifts), stellar masses, and specific star formation rates (sSFRs) obtained from the template-fitting method \textsc{LePHARE} \cite{Arnouts1999,Ilbert2006}. 
These parameters are canonical physical descriptors of galaxies; they capture a galaxy's intrinsic luminosity, current evolutionary state, and distance from Earth. To obtain image cutouts, we use the DAWN JWST Archive (DJA)\footnote{\url{https://dawn-cph.github.io}} and the image cutout tool from the HSC release website\footnote{\url{https://hsc-release.mtk.nao.ac.jp/doc/index.php/data-access__pdr3/}}.

For our MMU-crossmatched surveys we assemble three-channel RGB composites from the raw flux cubes: $g$, $r$, $z$ for HSC and Legacy Survey, and F090W, F277W, F444W for JWST, ensuring maximum wavelength coverage while remaining suitable for our foundation models trained on RGB natural images. 
Pixel intensities are normalised per band with an arcsinh stretch anchored to precomputed 1st and 99th percentile flux values (estimated per survey over batches of up to 10\,000 images):
$\bar{x} = \text{arcsinh}(\alpha (x - p_1) / (p_{99} - p_1))/\text{arcsinh}(\alpha)$, with stretch parameter $\alpha = 20$ and the output clipped to $[0, 1]$.
Arcsinh stretching preserves faint low-surface-brightness structure while compressing the dynamic range of bright galaxy centres, we choose $\alpha = 20$ following \citet{ref_lupton2004}. In the COSMOS-Web dataset, we use $g$, $r$, $z$ for HSC and $F115W$, $F150W$, and $F277W$ for JWST.
We crop HSC and Legacy Survey galaxies and bilinearly rescale the image cutouts to match the $3.84''\times3.84''$ angular size of the JWST cutouts ($96\times96$ pixels at $0.04''/\mathrm{pixel}$).
Finally, before passing data into a model, any model-specific preprocessing is applied according to each model's published processing pipeline.

\textbf{Model architectures.} Our model selection (Tab.~\ref{tab_models}) is designed to test the PRH along three axes.
First, we include models spanning supervised classification (ViT, ConvNeXtv2), self-supervised learning via knowledge distillation (DINOv3), joint-embedding prediction (IJEPA, VJEPA), and masked autoencoding (ViT-MAE), allowing us to test whether convergence depends on training objective. 
Second, we include vision-language models (CLIP, LLaVA-1.5, PaliGemma) to test whether architectures trained to bridge vision and language converge toward the same representations as unimodal vision models. 
Third, we include two astronomy-specific models: AstroPTv2, an autoregressive transformer pre-trained exclusively on DESI Legacy Survey imaging, and Specformer, a transformer trained on one-dimensional DESI spectra. 
AstroPTv2 tests whether domain-specific pre-training yields representations that diverge from or converge with those of general-purpose models.
Specformer represents our most out-of-distribution test case: a fundamentally different input modality (spectra vs. images) processed by a model that has never seen natural images.
We extract our embeddings from the literature-recommended embedding layer of each model, or from the penultimate layer where there is no recommendation.
For each architecture family, we test multiple model sizes to measure whether representational alignment scales with capacity or performance.

\begin{wraptable}{r}{0.5\textwidth}
\vspace{-1em}
\caption{Foundation models used in this study, spanning supervised, self-supervised, autoregressive, and multimodal training paradigms across vision, language, and spectral modalities.}
\vspace{0.5em}
\label{tab_models}
\footnotesize
\resizebox{\linewidth}{!}{%
\begin{tabular}{lll}
\toprule
Model & Training regime & Modality \\
\midrule
\multicolumn{3}{l}{\textit{Supervised}} \\
ViT \citep{ref_dosovitskiy2020} & Supervised classification & Images \\
ConvNeXtv2 \citep{ref_woo2023} & Supervised classification & Images \\
\addlinespace
\multicolumn{3}{l}{\textit{Self-supervised}} \\
DINOv3 \citep{2025arXiv250810104S} & Self-distillation & Images \\
IJEPA \citep{ref_assran2023} & Joint-embedding prediction & Images \\
VJEPA \citep{ref_assran2025} & Joint-embedding prediction & Video \\
ViT-MAE \citep{ref_he2022} & Masked autoencoding & Images \\
\addlinespace
\multicolumn{3}{l}{\textit{Multimodal / Vision-Language}} \\
LLaVA-1.5 \citep{ref_liu2023llava2} & Vision-language & Images + Text \\
PaliGemma \citep{ref_beyer2024} & Vision-language & Images + Text \\
CLIP \citep{ref_radford2021} & Contrastive image-text & Images + Text \\
    \addlinespace

\multicolumn{3}{l}{\textit{Astronomy-specific}} \\
AstroPTv2 \citep{ref_smith2024} & Next-token prediction & Astro.\ images \\
Specformer \citep{ref_parker2024} & Contrastive & Astro.\ spectra \\
\bottomrule
\end{tabular}
}
\end{wraptable}

\textbf{Metrics used to measure representational alignment.} Following the methodology established in the original PRH work, we measure representational alignment using the mutual $k$-nearest neighbour (MKNN) metric \citep{ref_chechik2010}. 
Given two embeddings $(\mathbf{z}_1, \mathbf{z}_2)$ corresponding to the same object as viewed by two different instruments or models, the MKNN score is computed as the cardinality of intersections for each object's $k$-nearest neighbours in the embedding space:
$
    \text{MKNN}(\mathbf{z}_1, \mathbf{z}_2) = k^{-1} | N_k(\mathbf{z}_{1}) \cap N_k(\mathbf{z}_{2})|,
$
averaged over all objects in the dataset, where $N_k$ is the $k$-nearest neighbours operation under the cosine distance, and $|\cdot|$ denotes set cardinality. 
We follow \citet{ref_huh2024} and set $k = 10$ across all experiments.
We complement MKNN with the Centered Kernel Alignment (CKA) metric \citep{ref_kornblith2019}:
$
    \text{CKA} = (\|A B^T\|^2_F) / (\|A A^T\|_F \|B B^T\|_F),
$
where $A$ and $B$ are mean centered embedding matrices and $\| \cdot \|_F$ denotes the Frobenius norm. We use both MKNN and CKA as MKNN captures local neighborhood agreement, whereas CKA captures global embedding similarity. To eliminate confounds from network scale we follow \citet{ref_groger2026} and use permutation-calibrated versions of the MKNN and CKA metrics in all of our experiments, which subtract an empirical null obtained by shuffling sample correspondences to correct for width-driven baseline inflation.

\section{Experiments and results}\label{sec_results}

We ask first whether foundation models encode galaxy physics (\S\ref{subsec_probes}), and then whether the geometry of their embeddings converges in ways that track this physical content (\S\ref{subsec_mknncka}).

\subsection{Physical interpretation experiments}\label{subsec_probes}

\begin{SCfigure}[][htbp]  
    \includegraphics[width=0.72\textwidth]{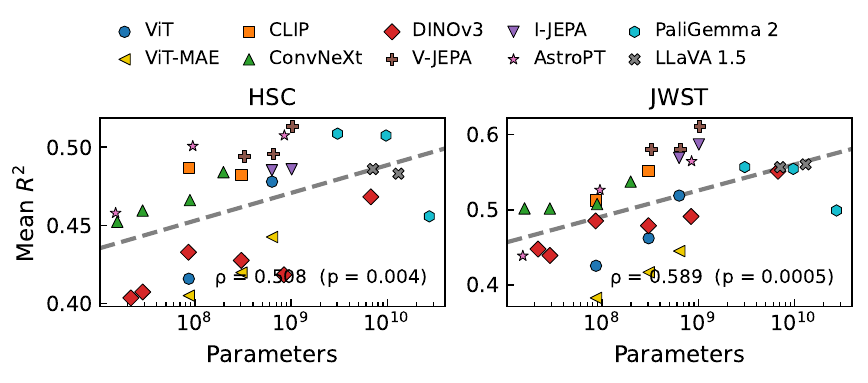}
    \caption{Linear probe $R^2$ as a function of model size, averaged over mass, sSFR, and redshift. 
    We find a significant positive correlation, even though AstroPT is the only model in our basket  pretrained substantially on galaxy imagery.}
    \label{fig_r2_scaling}
\end{SCfigure}

\textbf{Larger models encode more physics knowledge.}
We focus on three physical properties: stellar mass ($M_\star$), specific star formation rate (sSFR), and redshift ($z$). 
A model that has internalised the physics of galaxy images should encode all three canonical descriptors. 
After clipping values to the 1st and 99th percentiles to remove outliers, we train linear probes on the embeddings of each model to predict these properties. 
For both HSC and JWST, we run 10-fold cross-validation using the 45\,000 selected galaxies with matched physical parameter measurements from the COSMOS-Web catalog. 
Fig.~\ref{fig_r2_scaling} shows the average $R^2$ score across properties versus the parameter count with full per-property results reported in App.~\ref{app_physics}.
A Spearman's rank test gives a significant positive correlation for our JWST ($\rho=0.508, p=0.004$) and HSC ($\rho=0.589, p=0.0005$) cases, indicating that larger models recover more of the underlying astrophysics from images alone, despite AstroPT being our only tested model pre-trained significantly on galaxies (see  App.~\ref{app_pretraining_provenance}).
This relationship persists after controlling for architecture family in an ANCOVA, $R^2 \sim \log N_{\rm params} + C(\text{family})$, with a positive capacity coefficient for both HSC ($p=0.002$) and JWST ($p=2e{-5}$).
The scaling trend therefore cannot be explained by familial differences.

\vspace{-0.5em}

\begin{wrapfigure}[30]{r}{0.6\textwidth}
    \vspace{-1.5em}
    \includegraphics[width=0.6\textwidth]{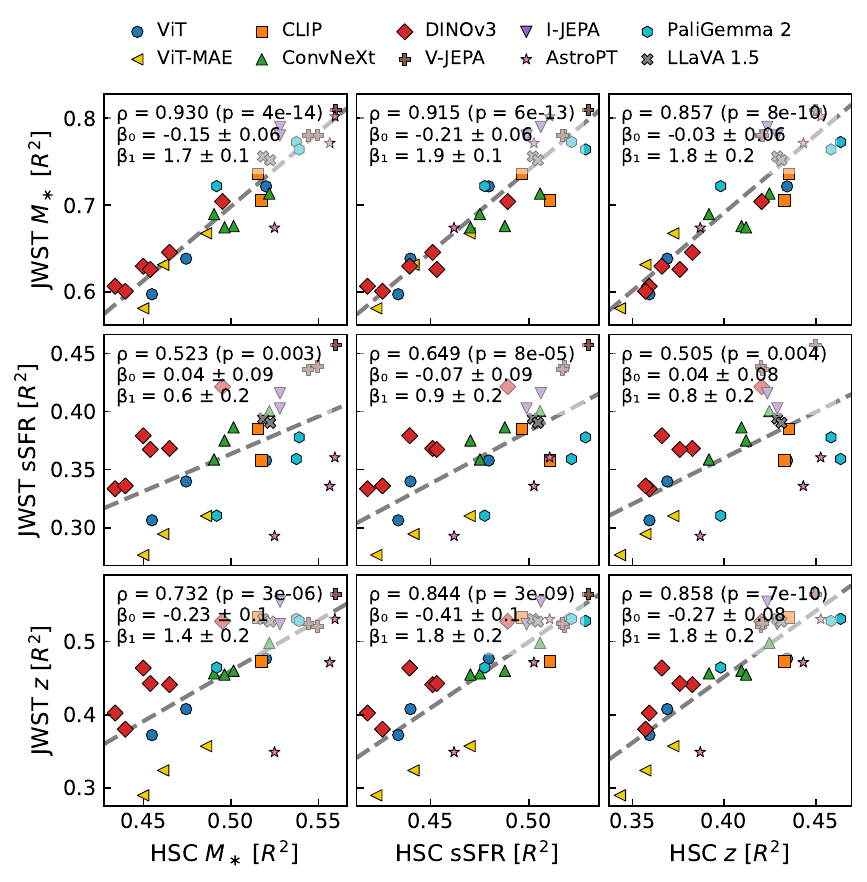}
    \caption{Linear probe $R^2$ scores of JWST versus HSC modalities on a grid of physical parameters. 
    Model performance is strongly correlated between the instruments, indicating that a model's capacity to linearly encode physical information is a property of its representations rather than a function of the observed wavelength range or modality.}
    \label{fig_jwst_vs_hsc_r2_grid}
\end{wrapfigure}

\textbf{Physics performance is correlated between modalities.}
In Fig. \ref{fig_jwst_vs_hsc_r2_grid}, we show JWST $R^2$ scores as a function of HSC $R^2$ scores in a grid of $M_\star$, sSFR, and $z$. We find a strong cross-modal correlation ($p{<<}0.01$ via a Spearman's rank test) in all cases, indicating that a model's capacity to linearly encode physical information is a consistent property of the model itself rather than that of the observed wavelength range.
A principal component analysis across the six probe $R^2$ scores confirms this with 87\% of the variance captured in a single axis with uniformly positive loadings---a `model quality' axis along which the same models that recover more astrophysics from optical imagery also recover more from near-infrared imagery, despite the two probing largely different physical processes (\S\ref{sec_methods} \P1).
One would expect this if our models are converging on representations of the underlying galaxies rather than on modality-specific image statistics.

JWST samples near-infrared stellar light, where emission is dominated by relatively low mass (long living) stars.
At near-IR wavelengths, the overall spectrum is also less affected by dust attenuation.
JWST imagery is therefore a clean tracer of already-built stellar mass but carries little  information about ongoing star formation.
HSC instead samples UV/optical light, and so is expected to be the more reliable tracer of recent and ongoing star formation \citep{ref_conroy2013}.
Finally, JWST's longer wavelength baseline pins down redshift more tightly than HSC's \citep{ref_salvato2019}, so we expect superunity linear regression slopes for $M_\star$ and $z$ on Fig.~\ref{fig_jwst_vs_hsc_r2_grid}'s diagonal and a subunity slope for sSFR---which is what we recover.
By definition, $\text{sSFR} \equiv \text{SFR}/M_\star$, so HSC sSFR carries information about stellar mass as well as star formation.
If both modalities encode mass in a shared, modality-invariant way, then the $M_\star$ component of HSC sSFR should align with JWST's mass representation. 
Because JWST carries little SFR information but traces $M_\star$ cleanly, we therefore expect HSC sSFR to correlate more strongly with JWST $M_\star$ than with JWST sSFR---and indeed we find $\rho(\text{HSC sSFR, JWST }M_\star) = 0.915$ vs $\rho(\text{HSC sSFR, JWST sSFR}) = 0.649$.
Taken holistically, our crossmodal correlations are consistent with our models having internalised modality invariant astrophysical content rather than instrument-specific image features, and foreshadows our further crossmodal analysis in \S\ref{subsec_mknncka}.

\begin{SCfigure}[][htbp] 
    \includegraphics[width=0.65\textwidth]{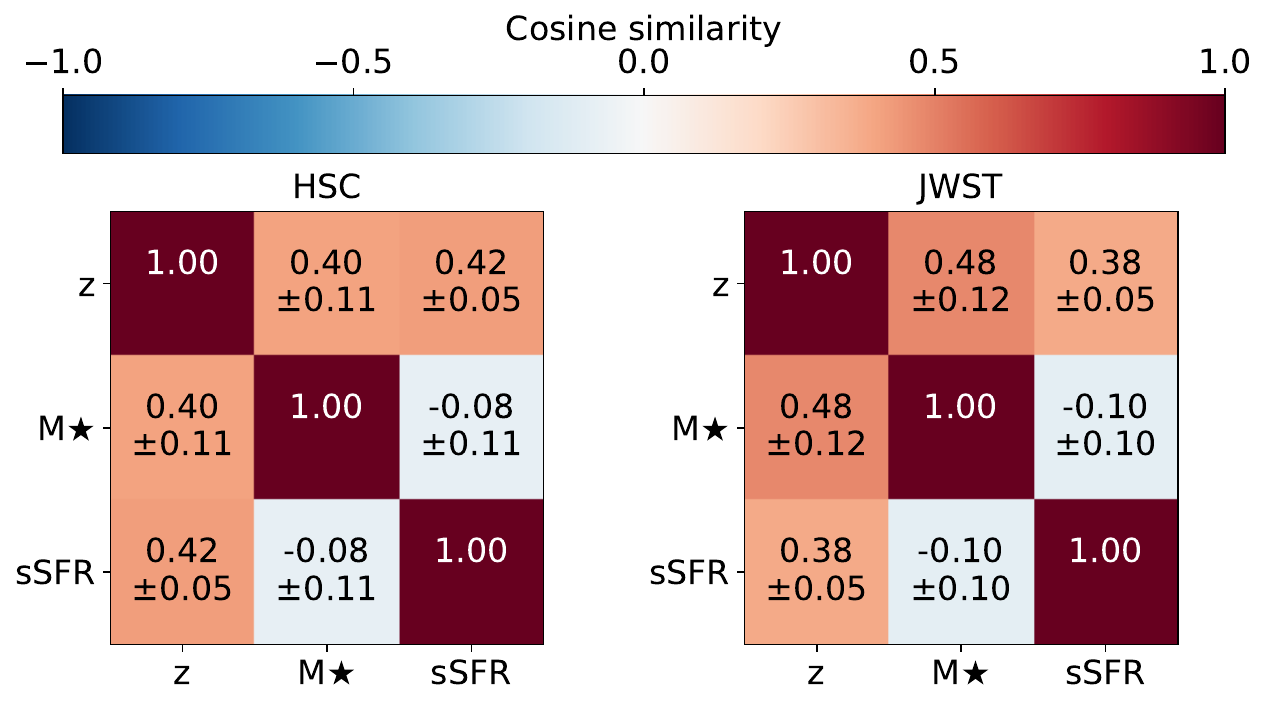}
    \caption{Cosine similarity matrices between the redshift ($z$), stellar mass ($M_\star$), and sSFR probe weight vectors for HSC (left) and JWST (right), averaged over all models. The relationships between the three probe weight vectors are consistent across all models, suggesting that the embedding spaces encode physics via similar internal organisation.}
    \label{fig_avg_cos_sim_matrix}
\end{SCfigure}

\textbf{Relative orientations of probe directions are consistent across models and with known physics.}
We assess the relative orientations of the redshift, stellar mass, and sSFR directions in embedding space by measuring their pairwise cosine similarities. 
Fig.~\ref{fig_avg_cos_sim_matrix} shows the cosine similarity matrices for HSC (left) and JWST (right), averaged across all models.
Within the star-forming population, galaxies follow the main sequence, $\log \text{SFR} = \alpha \log M_\star + \beta$, with a sub-linear slope $\alpha \lesssim 1$ \citep{2014ApJS..214...15S}.
As $\text{sSFR} \equiv \text{SFR}/M_\star$ it follows that $\log \text{sSFR} \propto (\alpha - 1) \log M_\star$, and so we would expect the sSFR-$M_\star$ correlation to be mildly negative.
This trend is reinforced by the rising quiescent (non-star forming) galaxy fraction with stellar mass, which shifts more massive galaxies to lower sSFR, and
both effects are weak relative to the several orders of magnitude that sSFR spans at fixed mass \citep{ref_wetzel2012}.
All of which is consistent with the small negative cosine recovered across all our tested models in Fig.~\ref{fig_avg_cos_sim_matrix}.
The correlation between $M_\star$ and redshift arises primarily from a Malmquist-style selection effect \citep{ref_malmquist1922}: since higher-redshift galaxies appear fainter, only intrinsically luminous---and therefore more massive---galaxies exceed the detection threshold of telescopes, biasing high-redshift samples toward larger stellar masses. 
Finally, higher redshift galaxies are observed to have higher star-forming rates \citep{2014ARA&A..52..415M}, which explains the sSFR and redshift correlation. 
In short, the correlations between these three parameters result from intrinsic galaxy physics and selection effects, yet they are captured by all models despite all-but-one of our models having been pre-trained on natural imagery or text (App.~\ref{app_pretraining_provenance}).

\vspace{-1em}

To summarize, we find that model capacity translates into physical knowledge, despite only one model in our basket being substantially pre-trained on galaxy imagery.
We also see that probe performance is tightly correlated between HSC and JWST across all three physical parameters, indicating that a model's ability to linearly encode galaxy physics transfers across observed wavelength ranges and observables.
And the relative orientations of the physical probe directions agree across models, and agreement reflects genuine astrophysical correlations, which suggests the formation of a shared physics subspace within the models' embeddings.

\subsection{Global and local embedding alignment experiments} \label{subsec_mknncka}

We test whether embedding alignment scales with model capacity and physics performance within model families, across modalities, and across architectures, using permutation-calibrated MKNN and CKA \citep[\S\ref{sec_methods} \P3;][]{ref_groger2026}.

\textbf{DINOv3 and PaliGemma reveal limits to convergence under scale.}
We first ask whether MKNN and CKA scores increase in the direction the PRH predicts.
We construct ordered pairs in two ways.
The first method takes consecutively sized pairs of pre-trained models, ordered by model size within a family (Tab.~\ref{tab_intramodal_app}); the PRH predicts that embedding similarity of pairs should increase with scale, since better-performing models sit closer to the Platonic Ideal.
The second method pairs a single model's embeddings of two distinct astronomical modalities (e.g.\ JWST and HSC imagery, Tab.~\ref{tab_crossmodal_app}); here we  expect better-performing models to produce more aligned cross-modal embeddings.
We treat each ordered pair as an independent Bernoulli trial with success if the similarity score increases along the predicted direction (larger model pair in the intra-architectural case, the larger model in the crossmodal case), and failure otherwise.
Under the null that scale carries no information about alignment, the success rate is 50\%, and we test the basket rate against this null via a binomial test.
Across our full basket, including DINOv3, the MKNN trend is in the predicted direction (60/96 = 62\%, $p = 0.02$), confirming the PRH's basic prediction. However, this effect is concentrated outside DINOv3, and excluding it sharpens the signal (51/69 = 74\%, $p = 9e{-5}$) while DINOv3 alone trends in the opposite direction (9/27 = 33\%). 
A Fisher exact test confirms that DINOv3 sits in a statistically distinct scaling regime from the rest of the basket ($p = 4e{-4}$ for MKNN, $p = 0.03$ for CKA).
CKA shows no significant trend in either case.

\begin{figure}[htb]
    \centering
    \includegraphics[width=\textwidth]{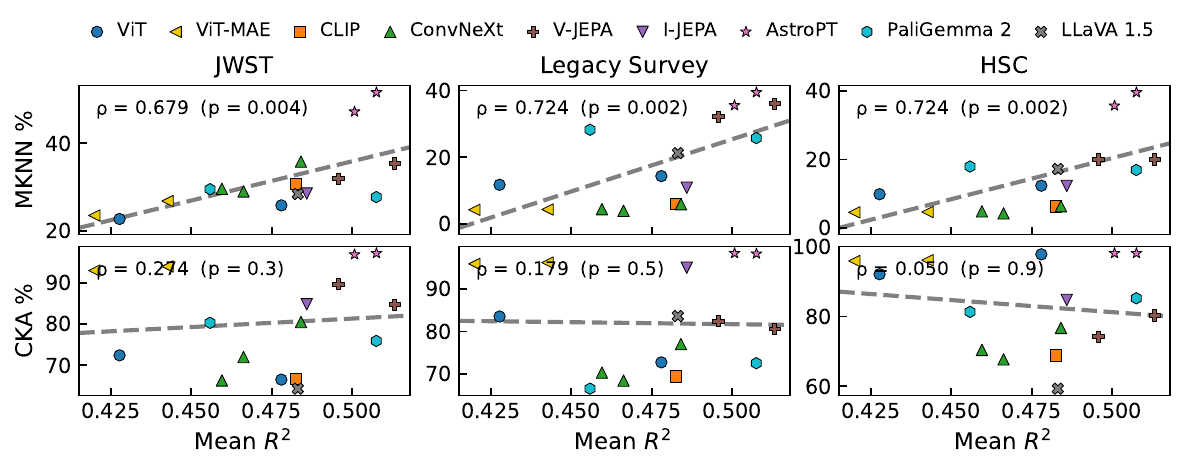}
    \caption{Intra-architectural scaling results.
    We plot our MKNN and CKA similarity values within each architecture family against the mean $R^2$ probe performance across the physical tasks described in \S\ref{subsec_probes}.
    In each pane we state the $\rho$ and $p$ values as measured via a Spearman's rank test.
    }
    \label{fig_intraarch_scaling}
\end{figure}

\begin{wrapfigure}[22]{r}{0.52\textwidth}
    \vspace{-3em}
    \includegraphics[width=0.52\textwidth]{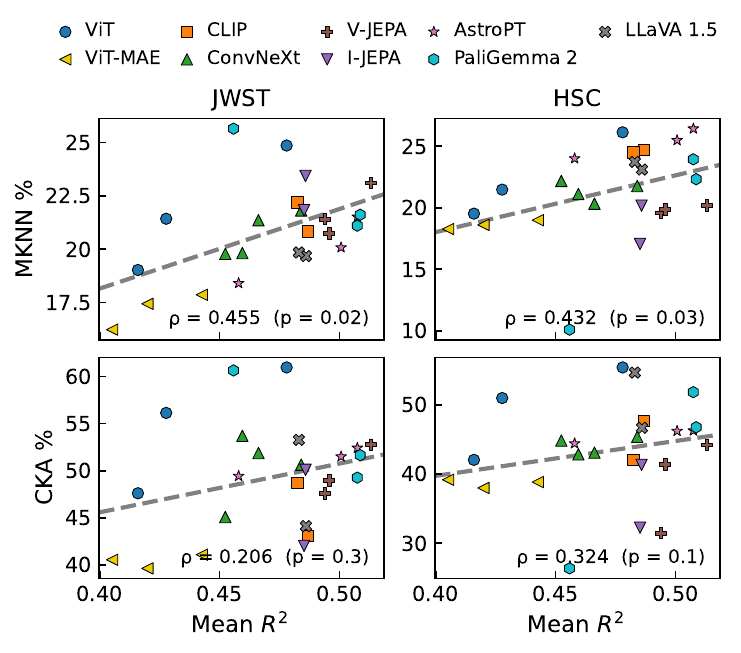}
    \caption{Cross-architecture metric distance (MKNN, CKA), plotted against average $R^2$ over the three physical properties for HSC and JWST images (\S\ref{subsec_probes}).
    We find that probe performance is significantly correlated with MKNN, but not with CKA.}
    \label{fig_crossarch_scaling}
\end{wrapfigure}

We investigate DINOv3's behavior by comparing its distilled students with their fixed ViT-7B/16 teacher and with the non-DINOv3 model basket.
On paired HSC and JWST imaging we find that teacher alignment as measured by MKNN generally increases with student size, while mean basket alignment decreases despite maintained or improving mean probe $R^2$ (App.~\ref{app_dinov3}).
This differs from the overall positive relationship between physical performance and basket alignment observed in our other models (Fig.~\ref{fig_crossarch_scaling}).
PaliGemma provides a complementary case, with crossmodal MKNN also trending negatively with scale. 
Its SigLIP vision encoder remains fixed in size, with additional capacity confined to the language trunk.
Layerwise probing shows that physical recovery declines through this trunk, and the largest model recovers less physics than its smaller counterparts (App.~\ref{app_paligemma}).
Taken together, these results suggest that convergence under scale depends on the training target and where capacity is added when scaling.
Given this statistically distinct scaling regime, we exclude DINOv3 from the remaining tests in this section.
We retain PaliGemma, as the same Fisher exact test does not reach significance ($p=0.74$ for MKNN).

\begin{figure}[htb]
    \centering
    \includegraphics[width=\textwidth]{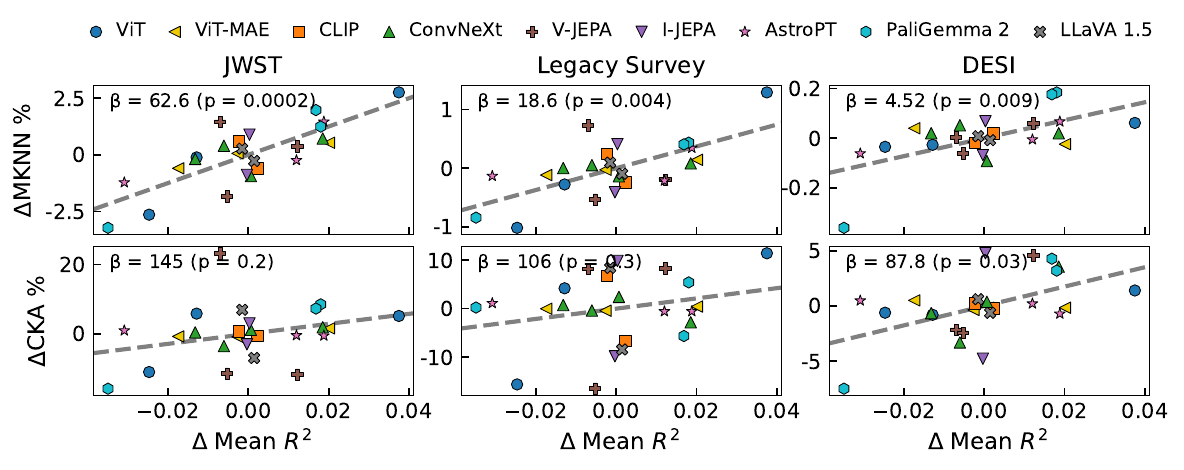}
\caption{
    Crossmodal MKNN and CKA metrics vs $R^2$ (\S\ref{subsec_probes}).
    $R^2$ and MKNN/CKA are demeaned per architecture family to show the trend without architecture family confounders.
    We state $\beta$ and $p$ from the ANCOVA tests $\text{MKNN} \sim R^2 + C(\text{family})$, and $\text{CKA} \sim R^2 + C(\text{family})$. 
}
    \label{fig_crossmodal_ancova}
\end{figure}

\textbf{Within-family local embedding alignment correlates with physics performance.}
We further test whether embedding similarities of models within the same architecture family tracks how well those models recover galaxy physics.
Concretely, for each model family we pair adjacent size rungs (e.g.\ ConvNeXtv2 Nano vs.\ Tiny, Tiny vs.\ Base, Base vs.\ Large), compute their MKNN and CKA scores on each of our three imaging surveys (JWST, Legacy, HSC; \S\ref{sec_methods} \P1), and compare these pairwise alignment scores against the physical-probe $R^2$ of the larger of the two models in the pair (\S\ref{subsec_probes}).
The PRH predicts that better-performing pairs should also be more internally aligned, since both models in a high-performing pair should sit closer to the Platonic Ideal.
Fig.~\ref{fig_intraarch_scaling} shows this is borne out for MKNN: across all three surveys we find a strong, significant Spearman correlation ($\rho=0.679$--$0.724$, $p<0.005$).
CKA shows no significant trend in any survey ($p \geq 0.3$).
Per-property breakdowns (mass, sSFR, redshift) and the full table of pairwise scores are reported in App.~\ref{app_intraarch}.

\textbf{Local embedding alignment across  modalities correlates with physics performance.}
We now test whether cross-modal embeddings of the same galaxies converge as model performance grows.
For each model we compute the MKNN and CKA alignment between the embeddings of HSC imagery and the embeddings of a second modality (JWST imagery, Legacy Survey imagery, or DESI spectra as processed by Specformer), and regress these alignment scores against probe $R^2$ (\S\ref{subsec_probes}).
In this test, if two instruments observing the same galaxies result in embeddings that align, the alignment cannot be inherited from raw pixel similarity and must instead reflect shared latent content recovered by the model.
Fig.~\ref{fig_crossmodal_ancova} summarizes our results.
Controlling for familial effects via an ANCOVA, we find a significant correlation ($p < 0.01$) in all tested MKNN cases (JWST: $\beta=62.6, p=2e-4$; Legacy: $\beta=18.6, p=4e{-3}$; DESI: $\beta=4.52, p=9e{-3}$), and no significant trend in the JWST and Legacy Survey CKA tests.
We find a borderline correlation in our CKA DESI test case ($p = 0.03$).
When additionally controlling for model capacity via $\mathrm{MKNN} \sim R^2 + \log_{10}N_{\mathrm{params}} + C(\mathrm{family})$ we find that the positive association between local alignment and physics performance persists for all three pairings (JWST: $\beta = 69.1$, $p = 4e{-4}$; Legacy: $\beta = 21.2$, $p = 5e{-3}$; DESI: $\beta = 6.79$, $p = 1e{-4}$). This indicates that parameter count does not explain the relationship, leaving the primary driver as probe performance.
Notably, we find a significant relation for DESI spectra vs HSC imagery: spectra and images share essentially no low-level statistics---a 1D sequence of fluxes vs a 2D RGB cutout, processed by a spectra model that has never seen an image, and image models that have never seen spectra---so alignment between them must be driven by shared physical content.
The finding that better-performing models produce more aligned crossmodal embeddings is direct evidence that convergence is mediated by underlying astrophysics rather than by input correlations or shared inductive biases inherited from natural-image pre-training. 
Per-property breakdowns are reported in App.~\ref{app_crossmodal}.

\textbf{Cross-architectural alignment correlates with physics performance.}
We ask whether our basket of models is converging toward a single `ideal' embedding space across architectures.
For each model in our basket we compute the MKNN and CKA scores against every other model and take the mean across comparisons, yielding a single per-model alignment score that quantifies how close a model sits to the rest of the basket in embedding space.
Fig.~\ref{fig_crossarch_scaling} plots the mean cross-architectural MKNN and CKA against the average linear probe $R^2$ as described in \S\ref{subsec_probes}. 
We restrict this analysis to the JWST-HSC crossmatched MMU dataset for computational tractability, since pairwise comparison across the full basket scales quadratically with model count.
We find a significant positive Spearman correlation between MKNN alignment and physics performance on both probe-spaces  ($\rho = 0.455, p=0.02$ for JWST, $\rho = 0.432, p = 0.03$ for HSC), indicating that models with better physics performance sit closer to the basket's average embedding geometry.
This trend holds across all tested physical properties (App.~\ref{app_crossarch}).
As in our earlier tests, we do not see a significant trend for CKA vs physics performance ($p > 0.1$ in both probe-spaces).

\textbf{Probe-defined physical directions contribute to embedding alignment.}
To complement our above analysis we remove the joint subspace spanned by our physical probe directions in 21 ViT--ConvNeXtv2 pairs evaluated on HSC imagery (App.~\ref{app_causal}).
We find that MKNN and CKA decrease in all cases, while not decreasing when removing a random subspace.
This supports the conclusion that the results reported here are caused by a learnt physical subspace in the models' embeddings.

Across all tests in this subsection we find that local embedding similarity (MKNN) tracks physics performance, while global similarity (CKA) does not.
The intra-architectural test shows that within a family, better-performing size-pairs are more locally aligned.
The crossmodal test extends this to pairs of instruments observing the same galaxies, and crucially holds for DESI spectra vs.\ HSC imagery---a comparison which isolates shared physical content as the driver of alignment.
The cross-architectural test shows that models that encode more physics sit closer to the basket's mean embedding geometry, suggesting convergence toward a shared representation rather than family-specific solutions.
The CKA null result across all four tests indicates that convergence is a property of local neighbourhood structure rather than global embedding geometry, consistent with the Aristotelian rather than the strict Platonic reading of the convergence hypothesis~\citep{ref_groger2026}.

\section{Conclusions: think local, not global}
\label{sec_discussion}

\textbf{We find strong support for the Aristotelian Representation Hypothesis.}
Across our tests in \S\ref{subsec_mknncka} we find consistently that local neighborhood structure (MKNN) tracks model performance, while global embedding similarity (CKA) does not.
We must remind ourselves that, aside from AstroPT and Specformer, our tested models are not significantly pre-trained on astronomical data (App.~\ref{app_pretraining_provenance}; that these models identify any correspondence between fundamentally different astronomical observations is remarkable, and suggests that sufficiently scaled up neural networks learn universal structural patterns transcending their training domains.
Our local-not-global embedding convergence pattern supports the Aristotelian reading of representational convergence \citep{ref_groger2026}: foundation models trained on different objectives, modalities, and data appear to agree on which galaxies are like other galaxies, without agreeing on a shared global coordinate system in which to place them.
We can also see that our natural image-trained models achieve embedding alignment that increases with model performance with Specformer's DESI spectral embeddings, therefore showing correspondence between fundamentally different modalities and data types they have never encountered that do not share low-level statistics.

\textbf{AstroPT does not dominate.}
AstroPTv2 is the only model in our basket pre-trained substantially on galaxy imagery, and yet it sits comfortably within the trends defined by general-purpose models rather than above them, in physics probe performance and in representational similarity.
We read this as evidence that sufficiently scaled general-purpose models already recover representations close to those delivered by domain-specific pre-training---precisely as Sutton's Bitter Lesson would predict \citep{ref_sutton2019}. 
This follows directly from the representational convergence documented throughout our remaining results.
Domain-specific architectures may still be useful where the input modality is genuinely outside the natural-image distribution (Specformer on 1D spectra is a clear example), but for galaxy imaging, scale and data diversity appear to substitute for domain specificity.

\textbf{DINOv3's and PaliGemma's anomalous behaviors are informative.}
Unlike our other tested models, DINOv3's distilled student models' embedding spaces converge towards their teacher model as their capacity increases rather than the remainder of the model basket.
PaliGemma's larger language trunks do not improve physical information recovery due to its fixed-capacity vision tower.
These results show that total parameter count is not sufficient to predict representational convergence while also providing guidance for pre-training and architectural choices for foundation models.

\textbf{Limitations.} 
While our results provide compelling evidence for the ARH, we note several limitations that suggest avenues for future exploration.
Our cross-modal samples are uneven in size (1.67k galaxies for JWST–HSC, 102k for Legacy–HSC, 18.6k for DESI–HSC).
We note however that our dataset sizes are in all cases larger than the datasets of size $n = 1024$ used in the establishing works of \citet[][Apps. C1, C2]{ref_huh2024} and \citet[][\S6.3]{ref_groger2026}.
We probe only three physical properties; convergence on morphology, kinematics, or environment may behave differently. 
Our cross-architectural test is restricted to our 1.67k JWST/HSC dataset for tractability.
The DINOv3 checkpoints confound distillation with other training differences, and isolating this effect requires a computationally-expensive controlled pre-training study.
We leave these questions for future studies.

\textbf{Implications for AI.}
This work demonstrates the usefulness of astronomy as an experimental instrument for understanding  machine learning models.
Open, crossmatched observations provide heterogeneous views of the same systems, while independently derived physical parameters provide an external reference.
We release the code, embeddings, and data so that the Platonic Universe can be reused to study neural representations beyond the hypotheses tested here.

\textbf{Implications for astronomical foundation modelling.} 
We observe general improvement in representational alignment at larger model scales, suggesting that each architecture is converging towards a shared representation.
Taken to its conclusion, this convergence implies that future efforts in astronomical foundation modelling should focus less on astronomy-specific architectures and more on scale and data diversity. It also follows that the astronomy community should embrace pre-trained foundation models rather than training from scratch:
if all architectures converge toward the same representations, then starting from models pre-trained on natural images or text---with their billions of parameters and massive computational investment already spent---offers both superior performance and dramatic reductions in environmental impact.
The broader open source machine learning community has already invested the GPU-centuries needed to learn general-purpose representations, we need now only gently guide these models toward astronomical use-cases.

\section*{AI use statement}

We used generative AI tools to aid data processing and coding efforts, checking any code changes via a manual PR review before merging into our repository.
In addition to this, we used generative AI tools to aid in editing and reviewing this manuscript after an initial human-written draft was in place.
We take responsibility for the final content of this manuscript, including text, claims, and any artifacts produced with the aid of generative AI.

\section*{Reproducibility statement}

In Tab.~\ref{tab_info} we list the datasets used, as well as the models used for this study.
For each dataset and model we provide a link to the publicly available data or weights.
Cross-matching between surveys is performed using MMU v1 with a 1" matching radius.

\begin{table}[ht]
    \centering
    \caption{Foundation models and astronomical datasets used in this study.}
    \label{tab_info}
    \small
    \resizebox{\textwidth}{!}{%
    \begin{tabular}{llll}
        \toprule
        Category & Name & Size & Hugging Face source \\
        \midrule

        Models
        & AstroPTv2  & 15M (Small)        & \href{https://huggingface.co/Smith42/astroPT_v2.0}{\texttt{Smith42/astroPT\_v2.0}} \\
        &            & 95M (Base)         & \href{https://huggingface.co/Smith42/astroPT_v2.0}{\texttt{Smith42/astroPT\_v2.0}} \\
        &            & 850M (Large)       & \href{https://huggingface.co/Smith42/astroPT_v2.0}{\texttt{Smith42/astroPT\_v2.0}} \\[5pt]

        & CLIP       & 86M (Base)         & \href{https://huggingface.co/openai/clip-vit-base-patch16}{\texttt{openai/clip-vit-base-patch16}} \\
        &            & 304M (Large)       & \href{https://huggingface.co/openai/clip-vit-large-patch14}{\texttt{openai/clip-vit-large-patch14}} \\[5pt]

        & ConvNeXtv2 & 15M (Nano)         & \href{https://huggingface.co/facebook/convnextv2-nano-22k-224}{\texttt{facebook/convnextv2-nano-22k-224}} \\
        &            & 28M (Tiny)         & \href{https://huggingface.co/facebook/convnextv2-tiny-22k-224}{\texttt{facebook/convnextv2-tiny-22k-224}} \\
        &            & 89M (Base)         & \href{https://huggingface.co/facebook/convnextv2-base-22k-224}{\texttt{facebook/convnextv2-base-22k-224}} \\
        &            & 198M (Large)       & \href{https://huggingface.co/facebook/convnextv2-large-22k-224}{\texttt{facebook/convnextv2-large-22k-224}} \\[5pt]

        & DINOv3     & 21M (ViT-S/16)     & \href{https://huggingface.co/facebook/dinov3-vits16-pretrain-lvd1689m}{\texttt{facebook/dinov3-vits16-pretrain-lvd1689m}} \\
        &            & 29M (ViT-S+/16)    & \href{https://huggingface.co/facebook/dinov3-vits16plus-pretrain-lvd1689m}{\texttt{facebook/dinov3-vits16plus-pretrain-lvd1689m}} \\
        &            & 86M (ViT-B/16)     & \href{https://huggingface.co/facebook/dinov3-vitb16-pretrain-lvd1689m}{\texttt{facebook/dinov3-vitb16-pretrain-lvd1689m}} \\
        &            & 300M (ViT-L/16)    & \href{https://huggingface.co/facebook/dinov3-vitl16-pretrain-lvd1689m}{\texttt{facebook/dinov3-vitl16-pretrain-lvd1689m}} \\
        &            & 840M (ViT-H+/16)   & \href{https://huggingface.co/facebook/dinov3-vith16plus-pretrain-lvd1689m}{\texttt{facebook/dinov3-vith16plus-pretrain-lvd1689m}} \\
        &            & 6.7B (ViT-7B/16)   & \href{https://huggingface.co/facebook/dinov3-vit7b16-pretrain-lvd1689m}{\texttt{facebook/dinov3-vit7b16-pretrain-lvd1689m}} \\[5pt]

        & IJEPA      & 632M (Huge)        & \href{https://huggingface.co/facebook/ijepa_vith14_22k}{\texttt{facebook/ijepa\_vith14\_22k}} \\
        &            & 1.0B (Giant)       & \href{https://huggingface.co/facebook/ijepa_vitg16_22k}{\texttt{facebook/ijepa\_vitg16\_22k}} \\[5pt]

        & LLaVA-1.5  & 7B                 & \href{https://huggingface.co/llava-hf/llava-1.5-7b-hf}{\texttt{llava-hf/llava-1.5-7b-hf}} \\
        &            & 13B                & \href{https://huggingface.co/llava-hf/llava-1.5-13b-hf}{\texttt{llava-hf/llava-1.5-13b-hf}} \\[5pt]

        & PaliGemma~2 & 3B                & \href{https://huggingface.co/google/paligemma2-3b-mix-224}{\texttt{google/paligemma2-3b-mix-224}} \\
        &             & 10B               & \href{https://huggingface.co/google/paligemma2-10b-mix-224}{\texttt{google/paligemma2-10b-mix-224}} \\
        &             & 28B               & \href{https://huggingface.co/google/paligemma2-28b-mix-224}{\texttt{google/paligemma2-28b-mix-224}} \\[5pt]

        & Specformer & 43M (Base)         & \href{https://huggingface.co/polymathic-ai/specformer}{\texttt{polymathic-ai/specformer}} \\[5pt]

        & ViT        & 86M (Base)         & \href{https://huggingface.co/google/vit-base-patch16-224-in21k}{\texttt{google/vit-base-patch16-224-in21k}} \\
        &            & 304M (Large)       & \href{https://huggingface.co/google/vit-large-patch16-224-in21k}{\texttt{google/vit-large-patch16-224-in21k}} \\
        &            & 632M (Huge)        & \href{https://huggingface.co/google/vit-huge-patch14-224-in21k}{\texttt{google/vit-huge-patch14-224-in21k}} \\[5pt]

        & ViT-MAE    & 86M (Base)         & \href{https://huggingface.co/facebook/vit-mae-base}{\texttt{facebook/vit-mae-base}} \\
&            & 304M (Large)       & \href{https://huggingface.co/facebook/vit-mae-large}{\texttt{facebook/vit-mae-large}} \\
        &            & 632M (Huge)        & \href{https://huggingface.co/facebook/vit-mae-huge}{\texttt{facebook/vit-mae-huge}} \\[5pt]

        & V-JEPA2    & 300M (ViT-L)       & \href{https://huggingface.co/facebook/vjepa2-vitl-fpc64-256}{\texttt{facebook/vjepa2-vitl-fpc64-256}} \\
        &            & 600M (ViT-H)       & \href{https://huggingface.co/facebook/vjepa2-vith-fpc64-256}{\texttt{facebook/vjepa2-vith-fpc64-256}} \\
        &            & 1.0B (ViT-G)       & \href{https://huggingface.co/facebook/vjepa2-vitg-fpc64-256}{\texttt{facebook/vjepa2-vitg-fpc64-256}} \\

        \midrule
        Crossmatches & JWST vs HSC      & 1.67k & \href{https://huggingface.co/datasets/Smith42/jwst_hsc_crossmatched}{\texttt{Smith42/jwst\_hsc\_crossmatched}} \\
                     & Lecy vs HSC    & 102k  & \href{https://huggingface.co/datasets/Smith42/legacysurvey_hsc_crossmatched}{\texttt{Smith42/legacysurvey\_hsc\_crossmatched}} \\
                     & DESI vs HSC      & 18.6k & \href{https://huggingface.co/datasets/Smith42/desi_hsc_crossmatched}{\texttt{Smith42/desi\_hsc\_crossmatched}} \\
        \midrule
        Embeddings   & & & \href{https://huggingface.co/datasets/UniverseTBD/pu-embeddings}{\texttt{UniverseTBD/pu-embeddings}}\\
        \bottomrule
    \end{tabular}
    }
\end{table}

We release all of our code on Github at \href{https://github.com/UniverseTBD/platonic-universe}{\texttt{github.com/UniverseTBD/platonic-universe}}.

\section*{Acknowledgements}

MJS would like to thank Regina Sarmiento, Rafael Mart\'inez-Galarza, Dun Li, and the AstroAI group for illuminating discussion and comments that improved this paper. 

We would like to thank Stella Biderman and EleutherAI for their compute and infrastructure support.

This research made use of the University of Hertfordshire's High Performance Computing facility (\url{https://uhhpc.herts.ac.uk/}).

This research used the DeltaAI advanced computing and data resource through allocation number PHY250286, which is supported by the National Science Foundation (award OAC 2320345) and the State of Illinois. DeltaAI is a joint effort of the University of Illinois Urbana-Champaign and its National Center for Supercomputing Applications.
Access to DeltaAI was granted through the Advanced Cyberinfrastructure Coordination Ecosystem: Services \& Support (ACCESS) program, which is supported by U.S. National Science Foundation grants \#2138259, \#2138286, \#2138307, \#2137603, and \#2138296.

This research used data obtained with the Dark Energy Spectroscopic Instrument (DESI). DESI construction and operations is managed by the Lawrence Berkeley National Laboratory. This material is based upon work supported by the U.S. Department of Energy, Office of Science, Office of High-Energy Physics, under Contract No. DE–AC02–05CH11231, and by the National Energy Research Scientific Computing Center, a DOE Office of Science User Facility under the same contract. Additional support for DESI was provided by the U.S. National Science Foundation (NSF), Division of Astronomical Sciences under Contract No. AST-0950945 to the NSF’s National Optical-Infrared Astronomy Research Laboratory; the Science and Technology Facilities Council of the United Kingdom; the Gordon and Betty Moore Foundation; the Heising-Simons Foundation; the French Alternative Energies and Atomic Energy Commission (CEA); the National Council of Science and Technology of Mexico (CONACYT); the Ministry of Science and Innovation of Spain (MICINN), and by the DESI Member Institutions: www.desi.lbl.gov/collaborating-institutions. The DESI collaboration is honored to be permitted to conduct scientific research on Iolkam Du’ag (Kitt Peak), a mountain with particular significance to the Tohono O’odham Nation. Any opinions, findings, and conclusions or recommendations expressed in this material are those of the author(s) and do not necessarily reflect the views of the U.S. National Science Foundation, the U.S. Department of Energy, or any of the listed funding agencies.

This work is based in part on observations made with the NASA/ESA/CSA James Webb Space Telescope. The data were obtained from the Mikulski Archive for Space Telescopes at the Space Telescope Science Institute, which is operated by the Association of Universities for Research in Astronomy, Inc., under NASA contract NAS 5-03127 for JWST.

The DESI Legacy Imaging Surveys consist of three individual and complementary projects: the Dark Energy Camera Legacy Survey (DECaLS), the Beijing-Arizona Sky Survey (BASS), and the Mayall z-band Legacy Survey (MzLS). DECaLS, BASS and MzLS together include data obtained, respectively, at the Blanco telescope, Cerro Tololo Inter-American Observatory, NSF’s NOIRLab; the Bok telescope, Steward Observatory, University of Arizona; and the Mayall telescope, Kitt Peak National Observatory, NOIRLab. NOIRLab is operated by the Association of Universities for Research in Astronomy (AURA) under a cooperative agreement with the National Science Foundation. Pipeline processing and analyses of the data were supported by NOIRLab and the Lawrence Berkeley National Laboratory (LBNL). Legacy Surveys also uses data products from the Near-Earth Object Wide-field Infrared Survey Explorer (NEOWISE), a project of the Jet Propulsion Laboratory/California Institute of Technology, funded by the National Aeronautics and Space Administration. Legacy Surveys was supported by: the Director, Office of Science, Office of High Energy Physics of the U.S. Department of Energy; the National Energy Research Scientific Computing Center, a DOE Office of Science User Facility; the U.S. National Science Foundation, Division of Astronomical Sciences; the National Astronomical Observatories of China, the Chinese Academy of Sciences and the Chinese National Natural Science Foundation. LBNL is managed by the Regents of the University of California under contract to the U.S. Department of Energy. The complete acknowledgments can be found at \url{https://www.legacysurvey.org/acknowledgment/}.

\section*{Contributors}

As a way to provide transparency in a large, multi-author scientific effort, we adopt the CRediT taxonomy\footnote{CRediT -- Contributor Roles Taxonomy: \url{https://credit.niso.org/}.} to report individual contributions. In this paper we define the contribution categories as:

\textbf{Conceptualization [C]:} Paper- or section-level intellectual framing: defining scope, research questions, experimental design, and narrative arc.

\textbf{Methodology [M]:} Designing the probe protocol, alignment metrics (MKNN, CKA), statistical tests, cross-matching, and preprocessing pipelines.

\textbf{Software [S]:} Implementing embedding extraction across model families, assembling cross-matched datasets, alignment metric code, probe training, statistical analyses, and figure production.

\textbf{Project administration [P]:} Soliciting and organizing inputs, managing timelines and revisions, integrating contributions, and ensuring internal consistency across contributors.

\textbf{Writing -- Original Draft [W]:} Substantive preparation of original text, including drafting new material and/or synthesizing multiple contributions into a coherent section or subsection.

\textbf{Writing -- Review \& Editing [R]:} Substantive review and revision of the manuscript text, including critical feedback, edits for clarity and correctness, and incorporation of reviewer comments.

Where shown, bracketed scope tags (e.g., [\S\ref{subsec_mknncka}]) indicate the section(s) associated with a listed role;
[all] denotes contributions spanning the full manuscript. Bolded scopes identify primary contributions,
and unbolded scopes identify secondary contributions.

\begin{center}
{\renewcommand{\arraystretch}{1.2}
\begin{tabular}{lp{0.75\textwidth}}
\toprule
Name & Contribution \\
\midrule
Trinidad Borrell & C[\S\ref{sec_methods} \P3, \S\ref{subsec_mknncka}, \S\ref{sec_discussion}], S[\S\ref{sec_methods} \P3, \S\ref{subsec_mknncka}], W[\S\ref{subsec_mknncka}], R[\S\ref{sec_methods}, \S\ref{sec_results},  \S\ref{sec_discussion}]\\
Steven Dillmann & C[\S\ref{sec_methods} \P3, \S\ref{subsec_mknncka}], M[\S\ref{sec_methods} \P3, \S\ref{subsec_probes}], S[\S\ref{sec_methods} \P3, \S\ref{subsec_probes}, \S\ref{subsec_mknncka}], R[\S\ref{sec_methods}, \S\ref{sec_results}]\\ 
Kshitij Duraphe & C[\S\ref{sec_methods} \P3, \S\ref{subsec_mknncka}], M[\S\ref{sec_methods} \P3, \S\ref{subsec_probes}], S[\textbf{\S\ref{sec_methods} \P3, \S\ref{subsec_probes}, \S\ref{subsec_mknncka}}, \S\ref{sec_methods} \P1], R[\S\ref{sec_methods}, \S\ref{sec_results}]\\ 
Furkan Eris & S[\S\ref{sec_methods} \P2, \S\ref{subsec_mknncka}], R[\S\ref{sec_methods}] \\
Kartheik Iyer &  R[\textbf{all}]\\
Ashod Khederlarian & C[\S\ref{sec_methods} \P1, \S\ref{subsec_probes}], M[\S\ref{sec_methods} \P1, \S\ref{subsec_probes}], S[\S\ref{sec_methods} \P1, \S\ref{subsec_probes}], W[\S\ref{sec_methods} \P1, \S\ref{subsec_probes}], R[\S\ref{sec_methods}, \S\ref{sec_results}]\\ 
Aman Kumar & C[\S\ref{sec_methods} \P3, \S\ref{subsec_mknncka}], S[\S\ref{sec_methods} \P2, \S\ref{sec_methods} \P3, \S\ref{subsec_mknncka}], W[\S\ref{subsec_mknncka}], R[\S\ref{sec_methods}] \\ 
Giovanni Marraffini & C[\S\ref{sec_methods} \P3,\S\ref{subsec_mknncka}], S[\S\ref{sec_methods} \P3, \S\ref{subsec_mknncka}], W[\S\ref{subsec_mknncka}], R[\S\ref{sec_methods}, \S\ref{sec_results}, \S\ref{sec_discussion}]\\
Michael J. Smith & C[\textbf{all}], M[\textbf{all}], S[\textbf{all}], P[\textbf{all}], W[\textbf{all}], R[\textbf{all}]\\
Shashwat Sourav & C[\S\ref{sec_methods} \P1,  \S\ref{sec_results}], M[\S\ref{sec_methods}, \S\ref{sec_methods} \P1, \S\ref{sec_methods} \P3, \S\ref{sec_results}], S[\S\ref{sec_methods} \P3], P[all], W[\S\ref{app_intraarch}], R[\S\ref{sec_methods}, \S\ref{sec_results}] \\
Rocco Di Tella & C[\S\ref{subsec_mknncka}, \S\ref{sec_discussion}],
S[\textbf{\S\ref{subsec_mknncka}, \S\ref{app_intraarch}}], R[\S\ref{sec_methods},
\S\ref{sec_results}, \S\ref{sec_discussion}]\\ 
John F. Wu & C[\S\ref{sec_methods} \P3, \S\ref{subsec_mknncka}], M[\S\ref{sec_methods} \P1,\S\ref{sec_methods} \P3], P[\textbf{all}], W[\S\ref{sec_methods}, \S\ref{sec_results}]\\
\bottomrule
\end{tabular}
}
\end{center}

\bibliography{main}
\bibliographystyle{iclr2027_conference}

\clearpage
\appendix

\section*{Contents}
\renewcommand{\contentsname}{\vspace{-2em}}  
\setcounter{tocdepth}{2}  
\tableofcontents

\clearpage

\section{Compute resources and experimental cost}
\label{sec_compute}

Extracting embeddings across the JWST, HSC, Legacy Survey, and DESI datasets cost the most GPU-hours, particularly for the larger vision-language models. In total, the experiments required approximately 3000 GPU-hours across various hardware tiers. Various downstream tasks (MKNN/CKA computation, linear probing, and statistical analyses) were computationally cheaper compared to extraction, requiring about one CPU-hour per probe once embeddings were cached. 

\begin{table}[htbp]
\centering
\caption{Hardware usage summary by chip type.}
\label{tab:compute_resources}
\begin{tabular}{lr}
\toprule
Hardware & Approximate Usage \\
\midrule
NVIDIA A40 / A80 & 300 GPU-hours \\
NVIDIA A100 / A800 & 600 GPU-hours \\
NVIDIA GH200 / H100 & 2000 GPU-hours \\
NVIDIA RTX 5070 and RTX 3090 & 50 GPU-hours \\
Standard CPU cores & $\approx 1$ CPU-hour per probe \\
\midrule
\textbf{Total} & \textbf{3,000 GPU-hours} \\
\bottomrule
\end{tabular}
\end{table}

\section{Dataset provenance}
\label{app_pretraining_provenance}

Here we discuss the datasets used to pretrain our examined models in order to rule out significant exposure to galaxy imagery.
ViT, ViT-MAE, ConvNeXtv2, and I-JEPA are pretrained on ImageNet-1k or ImageNet-22k \citep{ref_russakovsky2014}, neither of which includes a galaxy category. 
V-JEPA~2 uses ImageNet and video data.
These datasets provide no dedicated galaxy-image training, unlike AstroPTv2, which is explicitly pretrained on astronomical imagery. 
Incidental galaxy exposure remains possible in internet-derived corpora, including those used by CLIP, DINOv3, LLaVA, and PaliGemma~2.
We note that I-JEPA and V-JEPA~2 sit above the overall physics-scaling trend (Fig.~\ref{fig_r2_scaling}), including above AstroPTv2. 
Dedicated galaxy-image pretraining is therefore not necessary to recover the physical information measured by our probes.
We also note that we observe MKNN scaling between HSC imagery and DESI spectra (Fig.~\ref{fig_crossmodal_ancova}), with Specformer pretrained solely on spectra, further weakening an explanation based on shared image-level training examples.

\section{Extended physics results} \label{app_physics}

\begin{figure}[htbp]
    \centering
    \includegraphics[width=0.75\textwidth]{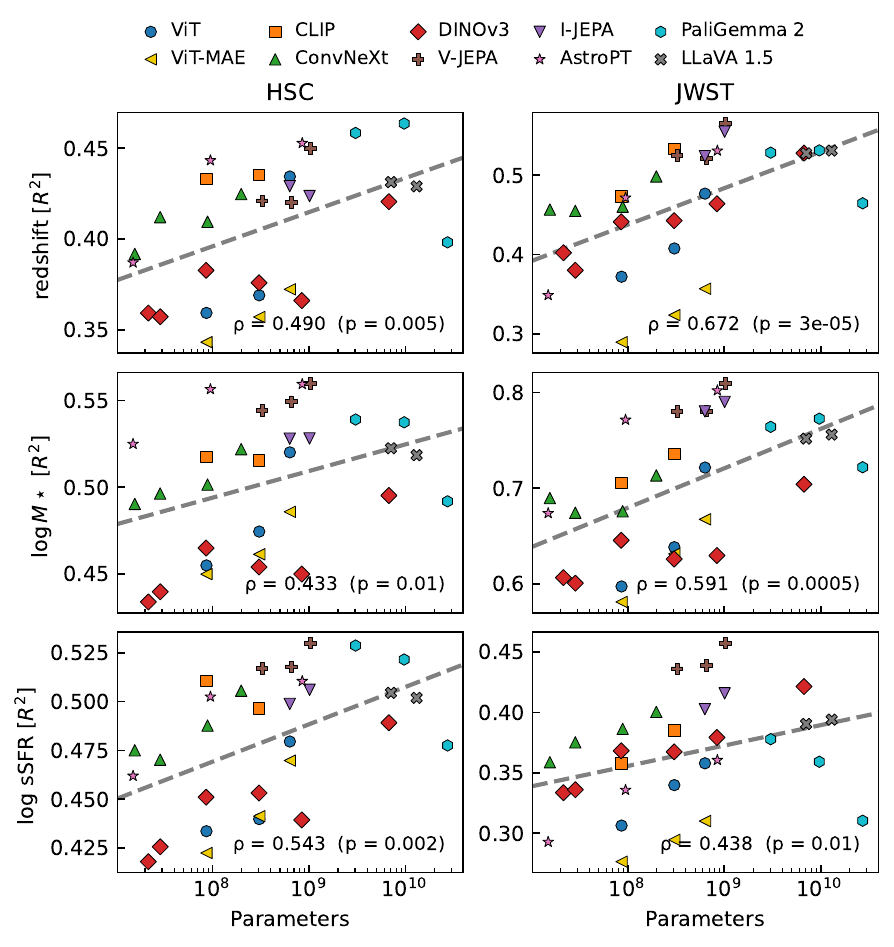}
    \caption{Parameter count vs probe performance for our tested physical properties. 
    Spearman's $\rho$ and $p$ values are stated for each panel.
    In all tested cases we find a significant correlation between parameter count and physics probe performance.}
    \label{fig_r2vsparams}
\end{figure}

\begin{figure}[p]
    \centering
    \includegraphics[width=1\textwidth]{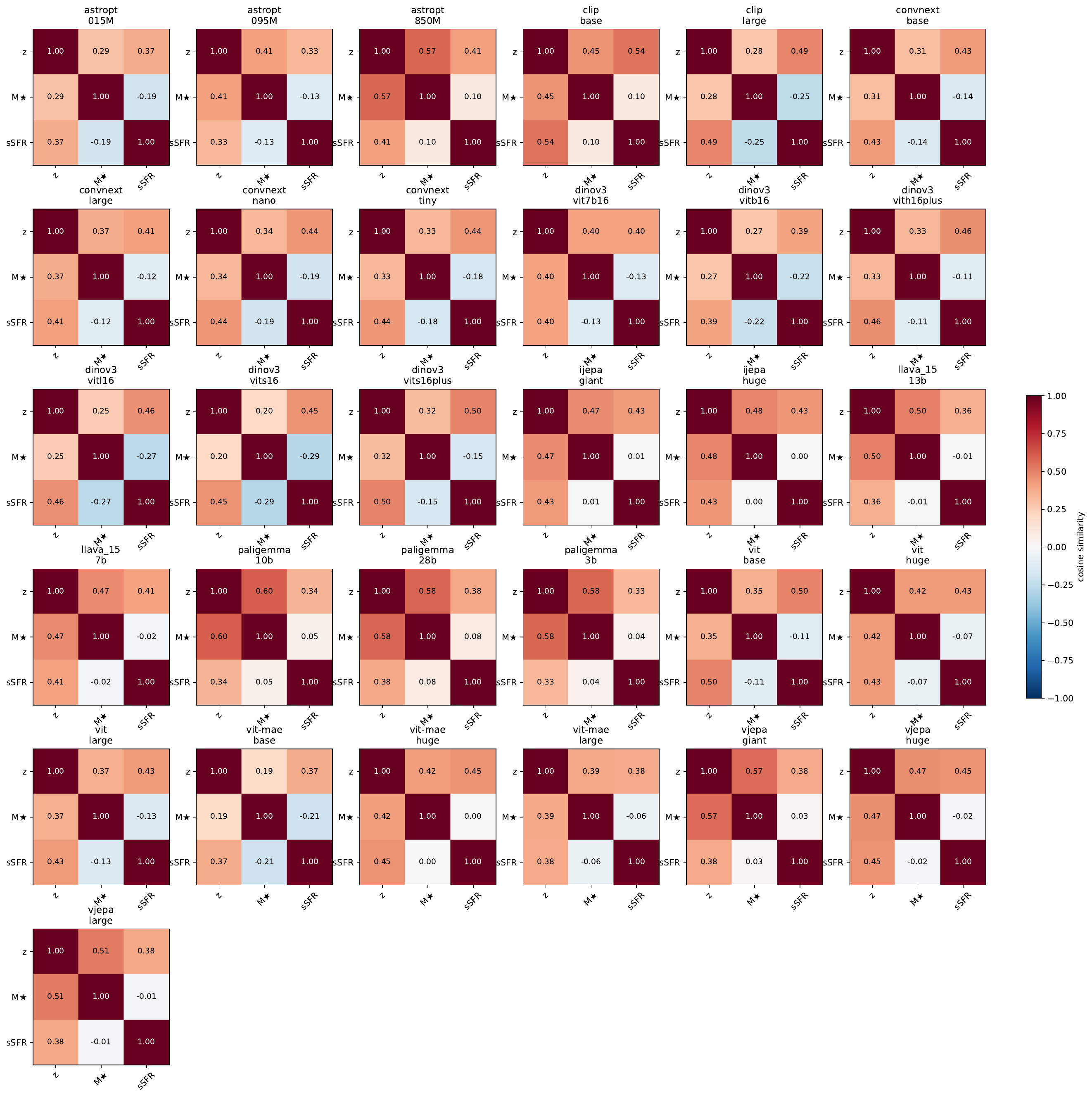}
    \caption{Cosine similarity matrices between the redshift ($z$), stellar mas ($M_\star$), and sSFR probe weight vectors using HSC images.}
    \label{fig_cos1}
\end{figure}

\begin{figure}[p]
    \centering
    \includegraphics[width=\textwidth]{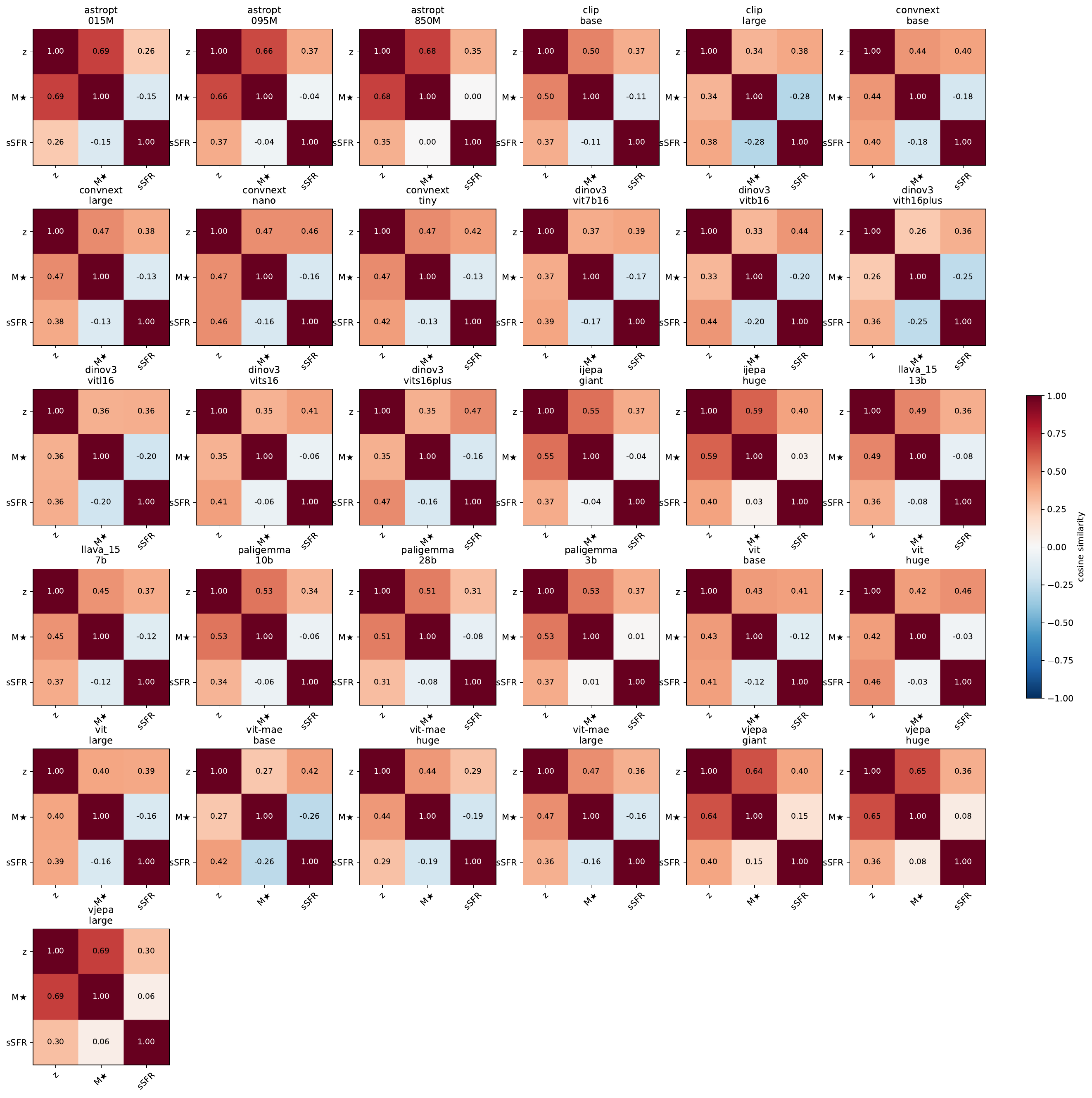}
    \caption{Cosine similarity matrices between the redshift ($z$), stellar mas ($M_\star$), and sSFR probe weight vectors using JWST images.}
    \label{fig_cos2}
\end{figure}

Here we show the full per-property results for the physics linear-probe experiments. Tab.~\ref{tab_r2_vs_params} lists mean and standard deviation $R^2$ values across 10-fold cross-validation for each model on HSC and JWST imagery, broken down by redshift, stellar mass, and sSFR.
Fig.~\ref{fig_r2vsparams} shows probe $R^2$ against parameter count separately for each of the three physical properties. 
Figs.~\ref{fig_cos1} and ~\ref{fig_cos2} show the per-model cosine similarity matrices between the redshift, stellar mass, and sSFR probe weight vectors for HSC and JWST imagery respectively.

\begin{table*}[htbp]
  \caption{
  Downstream regression performance ($R^2$) when predicting redshift ($z$), stellar mass ($\log M_\star$), and specific star formation rate (sSFR) from frozen embeddings of each model, evaluated on 45\,000 HSC and JWST galaxies.
  Values are mean $\pm$ standard deviation over 10 k-folds.
  }
  \label{tab_r2_vs_params}
  \centering
  \small
  \setlength{\tabcolsep}{4pt}
  \begin{tabular}{l ccc ccc}
    \toprule
    & \multicolumn{3}{c}{HSC ($R^2$)} & \multicolumn{3}{c}{JWST ($R^2$)} \\
    \cmidrule(lr){2-4} \cmidrule(lr){5-7}
    Model                         & $z$ & $\log M_\star$ & sSFR & $z$ & $\log M_\star$ & sSFR \\
    \midrule
    AstroPTv2 Small                 & $0.387_{\pm 0.013}$ & $0.525_{\pm 0.009}$ & $0.462_{\pm 0.002}$ & $0.349_{\pm 0.010}$ & $0.674_{\pm 0.010}$ & $0.293_{\pm 0.008}$ \\
    AstroPTv2 Base                 & $0.443_{\pm 0.017}$ & $0.556_{\pm 0.012}$ & $0.502_{\pm 0.006}$ & $0.472_{\pm 0.015}$ & $0.771_{\pm 0.008}$ & $0.336_{\pm 0.005}$ \\
    AstroPTv2 Large                & $0.453_{\pm 0.007}$ & $0.559_{\pm 0.013}$ & $0.510_{\pm 0.007}$ & $0.531_{\pm 0.012}$ & $0.802_{\pm 0.006}$ & $0.360_{\pm 0.005}$ \\
    \addlinespace
    ViT Base                      & $0.359_{\pm 0.006}$ & $0.455_{\pm 0.013}$ & $0.434_{\pm 0.006}$ & $0.372_{\pm 0.012}$ & $0.597_{\pm 0.004}$ & $0.306_{\pm 0.012}$ \\
    ViT Large                     & $0.369_{\pm 0.007}$ & $0.474_{\pm 0.007}$ & $0.440_{\pm 0.005}$ & $0.408_{\pm 0.010}$ & $0.638_{\pm 0.006}$ & $0.340_{\pm 0.009}$ \\
    ViT Huge                      & $0.434_{\pm 0.011}$ & $0.520_{\pm 0.008}$ & $0.480_{\pm 0.008}$ & $0.477_{\pm 0.008}$ & $0.722_{\pm 0.004}$ & $0.358_{\pm 0.009}$ \\
    \addlinespace
    ViT-MAE Base                  & $0.343_{\pm 0.010}$ & $0.450_{\pm 0.009}$ & $0.422_{\pm 0.008}$ & $0.290_{\pm 0.011}$ & $0.581_{\pm 0.009}$ & $0.277_{\pm 0.015}$ \\
    ViT-MAE Large                 & $0.357_{\pm 0.012}$ & $0.461_{\pm 0.009}$ & $0.441_{\pm 0.007}$ & $0.324_{\pm 0.012}$ & $0.631_{\pm 0.010}$ & $0.295_{\pm 0.011}$ \\
    ViT-MAE Huge                  & $0.372_{\pm 0.009}$ & $0.486_{\pm 0.008}$ & $0.470_{\pm 0.005}$ & $0.357_{\pm 0.011}$ & $0.668_{\pm 0.010}$ & $0.310_{\pm 0.010}$ \\
    \addlinespace
    DINOv3 vits16                 & $0.359_{\pm 0.005}$ & $0.434_{\pm 0.008}$ & $0.418_{\pm 0.008}$ & $0.402_{\pm 0.009}$ & $0.607_{\pm 0.006}$ & $0.334_{\pm 0.008}$ \\
    DINOv3 vits16plus             & $0.357_{\pm 0.008}$ & $0.440_{\pm 0.008}$ & $0.426_{\pm 0.008}$ & $0.380_{\pm 0.013}$ & $0.601_{\pm 0.008}$ & $0.336_{\pm 0.011}$ \\
    DINOv3 vitb16                 & $0.383_{\pm 0.005}$ & $0.465_{\pm 0.008}$ & $0.451_{\pm 0.011}$ & $0.441_{\pm 0.008}$ & $0.646_{\pm 0.007}$ & $0.368_{\pm 0.008}$ \\
    DINOv3 vitl16                 & $0.376_{\pm 0.004}$ & $0.454_{\pm 0.006}$ & $0.453_{\pm 0.008}$ & $0.443_{\pm 0.008}$ & $0.626_{\pm 0.006}$ & $0.367_{\pm 0.012}$ \\
    DINOv3 vith16plus             & $0.366_{\pm 0.010}$ & $0.450_{\pm 0.006}$ & $0.439_{\pm 0.008}$ & $0.464_{\pm 0.012}$ & $0.630_{\pm 0.007}$ & $0.379_{\pm 0.011}$ \\
    DINOv3 vit7b16                & $0.421_{\pm 0.007}$ & $0.495_{\pm 0.008}$ & $0.489_{\pm 0.013}$ & $0.528_{\pm 0.004}$ & $0.704_{\pm 0.003}$ & $0.421_{\pm 0.011}$ \\
    \addlinespace
    ConvNeXtv2 Nano               & $0.392_{\pm 0.005}$ & $0.490_{\pm 0.009}$ & $0.475_{\pm 0.007}$ & $0.457_{\pm 0.010}$ & $0.690_{\pm 0.007}$ & $0.359_{\pm 0.011}$ \\
    ConvNeXtv2 Tiny               & $0.412_{\pm 0.009}$ & $0.496_{\pm 0.009}$ & $0.470_{\pm 0.006}$ & $0.455_{\pm 0.016}$ & $0.675_{\pm 0.007}$ & $0.375_{\pm 0.008}$ \\
    ConvNeXtv2 Base               & $0.409_{\pm 0.004}$ & $0.501_{\pm 0.008}$ & $0.488_{\pm 0.011}$ & $0.460_{\pm 0.009}$ & $0.676_{\pm 0.003}$ & $0.386_{\pm 0.006}$ \\
    ConvNeXtv2 Large              & $0.425_{\pm 0.008}$ & $0.522_{\pm 0.010}$ & $0.506_{\pm 0.008}$ & $0.498_{\pm 0.012}$ & $0.713_{\pm 0.004}$ & $0.400_{\pm 0.010}$ \\
    \addlinespace
    I-JEPA Huge                   & $0.429_{\pm 0.008}$ & $0.528_{\pm 0.007}$ & $0.499_{\pm 0.006}$ & $0.524_{\pm 0.011}$ & $0.780_{\pm 0.006}$ & $0.403_{\pm 0.009}$ \\
    I-JEPA Giant                  & $0.424_{\pm 0.004}$ & $0.528_{\pm 0.007}$ & $0.506_{\pm 0.012}$ & $0.555_{\pm 0.013}$ & $0.790_{\pm 0.006}$ & $0.416_{\pm 0.008}$ \\
    \addlinespace
    V-JEPA2 Large                 & $0.421_{\pm 0.008}$ & $0.544_{\pm 0.010}$ & $0.517_{\pm 0.006}$ & $0.525_{\pm 0.010}$ & $0.780_{\pm 0.006}$ & $0.436_{\pm 0.007}$ \\
    V-JEPA2 Huge                  & $0.420_{\pm 0.006}$ & $0.549_{\pm 0.012}$ & $0.518_{\pm 0.006}$ & $0.521_{\pm 0.011}$ & $0.781_{\pm 0.005}$ & $0.439_{\pm 0.012}$ \\
    V-JEPA2 Giant                 & $0.450_{\pm 0.006}$ & $0.560_{\pm 0.011}$ & $0.530_{\pm 0.007}$ & $0.565_{\pm 0.011}$ & $0.810_{\pm 0.003}$ & $0.457_{\pm 0.009}$ \\
    \addlinespace
    CLIP Base                     & $0.433_{\pm 0.009}$ & $0.517_{\pm 0.008}$ & $0.511_{\pm 0.007}$ & $0.473_{\pm 0.009}$ & $0.706_{\pm 0.003}$ & $0.358_{\pm 0.011}$ \\
    CLIP Large                    & $0.435_{\pm 0.008}$ & $0.515_{\pm 0.010}$ & $0.496_{\pm 0.007}$ & $0.533_{\pm 0.009}$ & $0.736_{\pm 0.004}$ & $0.385_{\pm 0.010}$ \\
    \addlinespace
    PaliGemma 2 3B                & $0.458_{\pm 0.007}$ & $0.539_{\pm 0.008}$ & $0.529_{\pm 0.011}$ & $0.528_{\pm 0.011}$ & $0.764_{\pm 0.005}$ & $0.378_{\pm 0.007}$ \\
    PaliGemma 2 10B               & $0.464_{\pm 0.009}$ & $0.537_{\pm 0.011}$ & $0.522_{\pm 0.011}$ & $0.531_{\pm 0.015}$ & $0.773_{\pm 0.004}$ & $0.359_{\pm 0.008}$ \\
    PaliGemma 2 28B               & $0.398_{\pm 0.009}$ & $0.492_{\pm 0.009}$ & $0.478_{\pm 0.010}$ & $0.465_{\pm 0.014}$ & $0.722_{\pm 0.006}$ & $0.310_{\pm 0.014}$ \\
    \addlinespace
    LLaVA 1.5 7B                  & $0.431_{\pm 0.009}$ & $0.522_{\pm 0.013}$ & $0.504_{\pm 0.006}$ & $0.528_{\pm 0.014}$ & $0.752_{\pm 0.005}$ & $0.390_{\pm 0.010}$ \\
    LLaVA 1.5 13B                 & $0.429_{\pm 0.008}$ & $0.518_{\pm 0.009}$ & $0.502_{\pm 0.004}$ & $0.531_{\pm 0.011}$ & $0.756_{\pm 0.003}$ & $0.394_{\pm 0.012}$ \\
    \bottomrule
  \end{tabular}
\end{table*}

\subsection{Embedding width and probe capacity}
\label{app_probe_width}

Wider embeddings give a linear probe more trainable coefficients, raising the possibility that differences in physics-probe performance reflect probe capacity. To assess this possibility, we group 19 models by embedding width (384, 768, 1024, and 1280) and perform a one-way ANOVA comparing physics-probe performance across these groups, separately for HSC and JWST. 
We find no statistically significant differences between width groups: $F(3,15)=0.92$, $p=0.46$ for HSC and $F(3,15)=1.29$, $p=0.31$ for JWST.

\clearpage

\section{Anomaly investigation} \label{app_anomaly}

Here we provide extended analysis on DINOv3's and PaliGemma's anomalous behavior.

\subsection{DINOv3} \label{app_dinov3}

DINOv3's VIT-S/16 through ViT-H+/16 models are distilled from a fixed ViT-7B/16 teacher trained from scratch.
On the 1.67k galaxies with paired HSC and JWST imaging we compute the student models' MKNN alignment with the teacher and its mean pairwise alignment of the non-DINOv3 model basket.
Tab.~\ref{tab_dinov3_teacher_basket} reports the results.
We see that teacher alignment generally increases with student model size while basket alignment decreases.
This is unlike our other tested models where basket alignment tracks model performance (\S\ref{subsec_mknncka} \P4).
This pattern supports the interpretation that increasing student capacity improves alignment to a teacher model without necessarily improving alignment with independently trained models.

\begin{table*}[htbp]
  \caption{
  DINOv3 MKNN alignment with the ViT-7B/16 teacher and the non-DINOv3
  model basket, evaluated on 1.67k galaxies with paired HSC and JWST
  imaging. Basket alignment is averaged over non-DINOv3 models.
  Scores are percentages, with uncertainties estimated from 200 draws
  containing 80\% of the sample each.
  }
  \label{tab_dinov3_teacher_basket}
  \centering
  \small
  \setlength{\tabcolsep}{4pt}
  \begin{tabular}{l r cc cc}
    \toprule
    & & \multicolumn{2}{c}{HSC MKNN (\%)}
      & \multicolumn{2}{c}{JWST MKNN (\%)} \\
    \cmidrule(lr){3-4} \cmidrule(lr){5-6}
    Model & Params (M) & Teacher & Basket & Teacher & Basket \\
    \midrule
    ViT-S/16  & 21  & $25.0 \pm 0.3$ & $24.3 \pm 0.2$ & $27.8 \pm 0.3$ & $23.0 \pm 0.2$ \\
    ViT-S+/16 & 29  & $25.7 \pm 0.3$ & $23.2 \pm 0.2$ & $27.8 \pm 0.3$ & $21.5 \pm 0.2$ \\
    ViT-B/16  & 86  & $26.0 \pm 0.3$ & $23.1 \pm 0.2$ & $31.3 \pm 0.3$ & $22.4 \pm 0.2$ \\
    ViT-L/16  & 300 & $29.4 \pm 0.3$ & $18.7 \pm 0.2$ & $37.4 \pm 0.4$ & $18.0 \pm 0.2$ \\
    ViT-H+/16 & 840 & $29.3 \pm 0.3$ & $15.0 \pm 0.2$ & $40.5 \pm 0.3$ & $16.0 \pm 0.2$ \\
    \midrule
    ViT-7B/16 (teacher) & 6716
      & --- & $17.6 \pm 0.2$ & --- & $19.4 \pm 0.2$ \\
    \bottomrule
  \end{tabular}
\end{table*}

\subsection{PaliGemma} \label{app_paligemma}

\begin{figure}[htbp]
    \includegraphics[width=\textwidth]{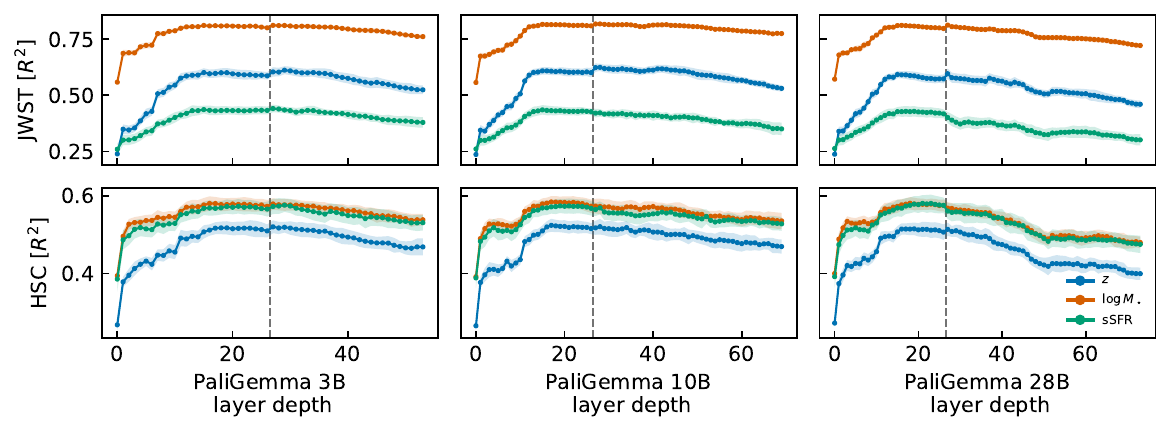}
    \caption{Physics linear $R^2$ probing results for the PaliGemma 3B, 10B, and 28B networks on a layer-wise basis.
    The $k=10$-fold standard deviation is shaded. 
    We demarcate the boundary between the fixed-size SigLIP image encoder and Gemma decoder networks with a gray dashed line.
}
\label{fig_paligemma_layers}
\end{figure}

Here we investigate PaliGemma's physics performance in more depth.
We run the linear probing tests described in \S\ref{subsec_probes} over each layer in the PaliGemma model of families.
Fig.~\ref{fig_paligemma_layers} shows our results.
We can see that the fixed-size SigLIP image encoder network eliminates the correlation between network size and physics recovery that we observe in other networks (Fig.~\ref{fig_r2_scaling}); we even find that the scaling of the Gemma LLM backbone network is detrimental to physics recovery!
This is likely due to nuisance noise being introduced by the language trunk to the astronomical signal passed by SigLIP, with more noise added the further the signal propagates through the Gemma trunk.

\clearpage

\section{Extended intra-architectural results} \label{app_intraarch}

\begin{figure}[htbp]
    \begin{subfigure}{1\textwidth}
        \includegraphics[width=\textwidth]{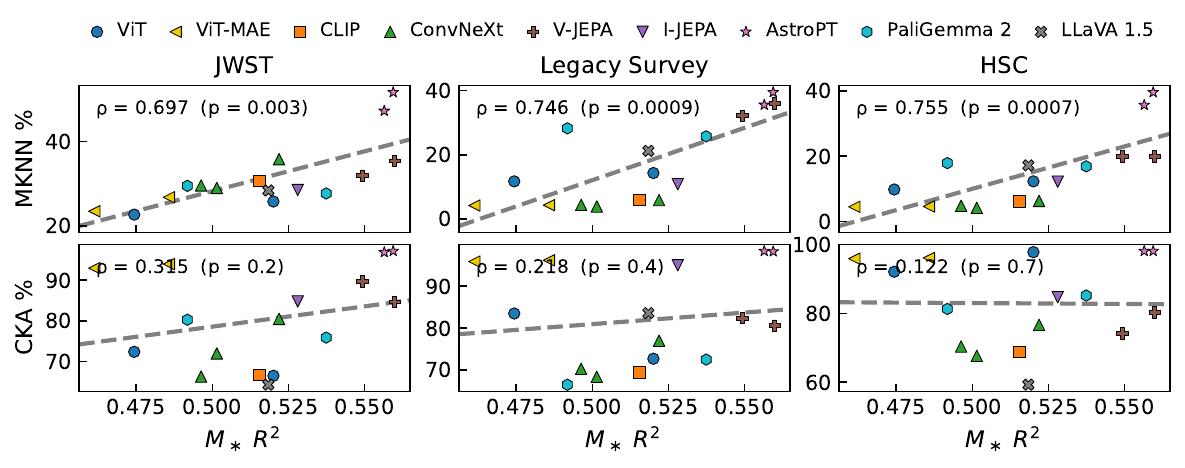}
    \caption{Mass $(M_\star)$.}
    \end{subfigure}
    \begin{subfigure}{1\textwidth}
        \includegraphics[width=\textwidth]{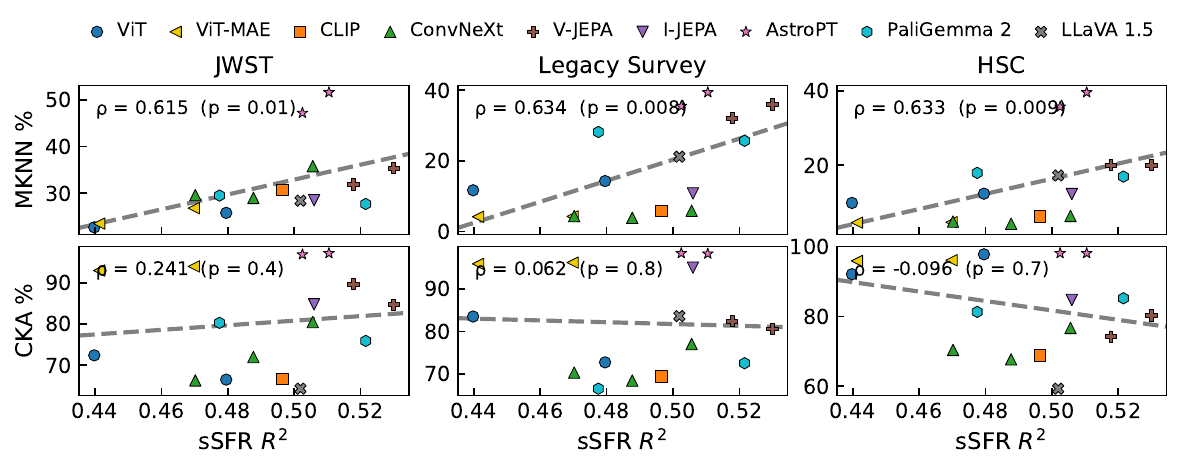}
    \caption{Specific Star Formation Rate (sSFR).}
    \end{subfigure}
    \begin{subfigure}{1\textwidth}
        \includegraphics[width=\textwidth]{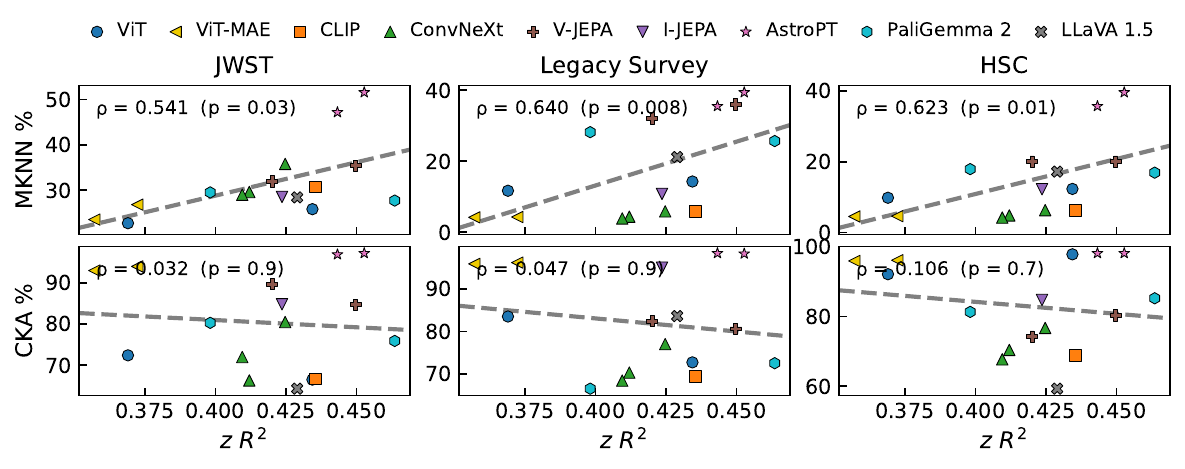}
    \caption{Redshift $(z)$.}
    \end{subfigure}
    \caption{Intra-architectural embedding alignment (MKNN and CKA) vs probe performance for our tested physical properties.
    Spearman's $\rho$ and $p$ values are stated for each panel.
    In all tested cases we find a significant correlation between MKNN and physics probe performance.
    We do not find a significant correlation between CKA and physics probe performance in any tested case.
}
\label{fig_intramodal_app}
\end{figure}

We list the full results across pairs of models within the same family in Tab.~\ref{tab_intramodal_app}. 
Each entry is the MKNN or CKA score as a percentage. 
Fig.~\ref{fig_intramodal_app} plots intra-architectural MKNN and CKA against linear-probe $R^2$ for each of redshift, stellar mass, and sSFR individually.

\begin{table*}[htbp]
  \caption{
  Intra-architectural embedding alignment within a model family, measured by MKNN and CKA scores.
  The PRH predicts that both MKNN and CKA scores will increase as we compare the embeddings of larger model pairs within a model family, since larger models will generate embeddings closer to the Platonic ideal representation.
  \vspace{0.5em}
  }
  \label{tab_intramodal_app}
  \centering
  \small
  \setlength{\tabcolsep}{6pt}
  \begin{tabular}{l ccc ccc}
    \toprule
    & \multicolumn{3}{c}{MKNN (\%)} & \multicolumn{3}{c}{CKA (\%)} \\
    \cmidrule(lr){2-4} \cmidrule(lr){5-7}
    Model Pairs & JWST & Legacy & HSC & JWST & Legacy & HSC \\
    \midrule
    AstroPTv2 Small vs Base      & 47.2 & 35.5 & 35.7 & 96.9 & 98.4 & 98.1 \\
    AstroPTv2 Base vs Large      & 51.6 & 39.4 & 39.6 & 97.2 & 98.3 & 98.1 \\
    \addlinespace
    CLIP Base vs Large           & 30.7 & 5.9  & 6.2 & 66.7 & 69.4 & 68.8 \\
    \addlinespace
    ConvNeXtv2 Nano vs Tiny      & 29.6 &  4.4 &  4.8 & 66.3 & 70.3 & 70.4 \\
    ConvNeXtv2 Tiny vs Base      & 29.0 &  3.9 &  4.2 & 72.0 & 68.4 & 67.7 \\
    ConvNeXtv2 Base vs Large     & 35.8 &  5.9 &  6.3 & 80.5 & 77.0 & 76.7 \\
    \addlinespace
    DINOv3 vits16 vs vits16plus  & 43.6 & 10.8 & 16.5 & 88.5 & 89.7 & 86.7\\
    DINOv3 vits16plus vs vitb16  & 38.2 &  9.6 & 14.3 & 86.5 & 88.5 & 86.3 \\
    DINOv3 vitb16 vs vitl16      & 32.7 &  7.1 &  9.4 & 73.6 & 73.1 & 68.0 \\
    DINOv3 vitl16 vs vith16plus  & 34.0 &  6.3 &  8.8 & 78.5 & 68.7 & 66.6 \\
    DINOv3 vith16plus vs vit7b16 & 40.5 &  7.8 & 11.7 & 81.8 & 62.1 & 64.2 \\
    \addlinespace
    IJEPA Huge vs Giant & 28.5 & 10.8 & 12.2 & 84.8 & 95.0 & 84.7 \\
    \addlinespace
    LLaVA-1.5 7B vs 13B & 28.4 & 21.2 & 17.2 & 64.3 & 83.6 & 59.3 \\
    \addlinespace
    ViT-MAE Base vs Large        & 23.5 &  4.2 &  4.5 & 93.0 & 95.9 & 95.9 \\
    ViT-MAE Large vs Huge        & 26.8 &  4.3 &  4.6 & 94.0 & 96.2 & 96.1 \\
    \addlinespace
    ViT Base vs Large           & 22.7 & 11.7 & 9.8 & 72.4 & 83.5  & 92.1 \\
    ViT Large vs Huge           & 25.8 & 14.3 & 12.3 & 66.5 & 72.7 & 97.8 \\
    \addlinespace
    PaliGemma 2 3B vs 10B        & 27.7 & 25.7 & 16.9 & 75.9 & 72.5  & 85.2 \\
    PaliGemma 2 10B vs 28B       & 29.5 & 28.2 & 17.9 & 80.3 & 66.5 & 81.3 \\
    \addlinespace
    V-JEPA2 ViT-L vs H           & 31.9 & 32.1 & 20.0 & 89.7 & 82.4 & 74.2 \\
    V-JEPA2 ViT-H vs G           & 35.4 & 36.0 & 20.0 & 84.7 & 80.6 & 80.3 \\
    \bottomrule
  \end{tabular}
\end{table*}

\clearpage

\section{Extended crossmodal results}
\label{app_crossmodal}

\begin{figure}[htbp]
    \begin{subfigure}{1\textwidth}
        \includegraphics[width=\textwidth]{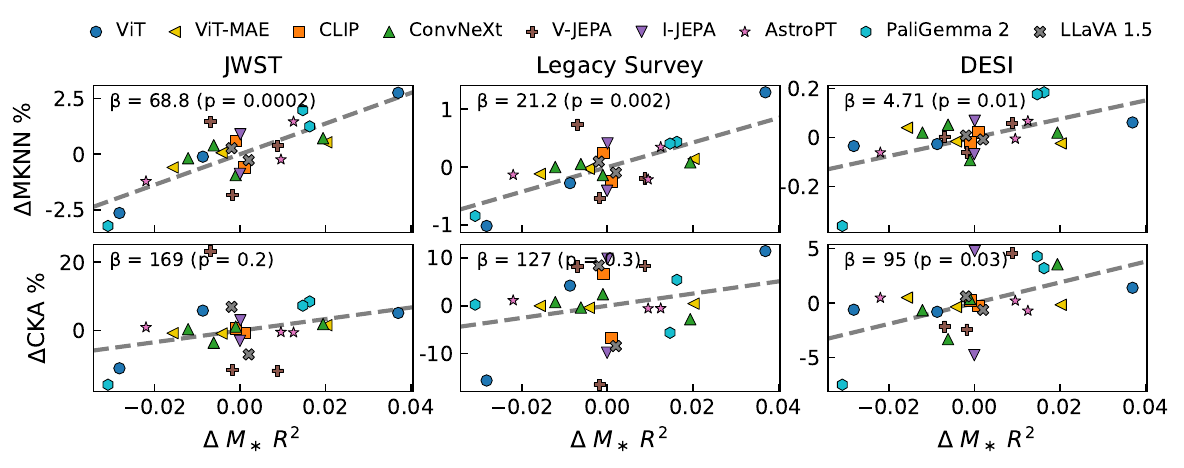}
    \caption{Mass $(M_\star)$.}
    \end{subfigure}
    \begin{subfigure}{1\textwidth}
        \includegraphics[width=\textwidth]{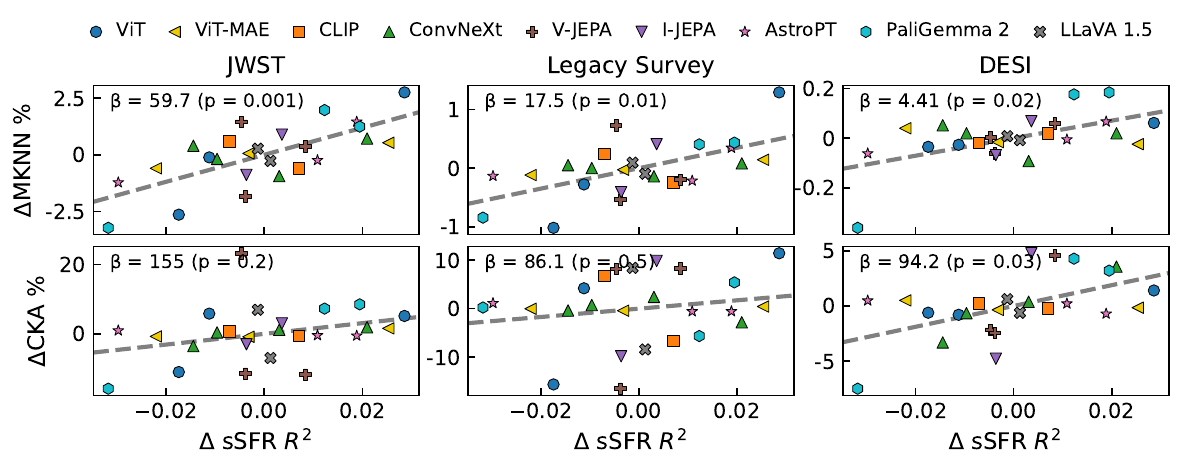}
    \caption{Specific Star Formation Rate (sSFR).}
    \end{subfigure}
    \begin{subfigure}{1\textwidth}
        \includegraphics[width=\textwidth]{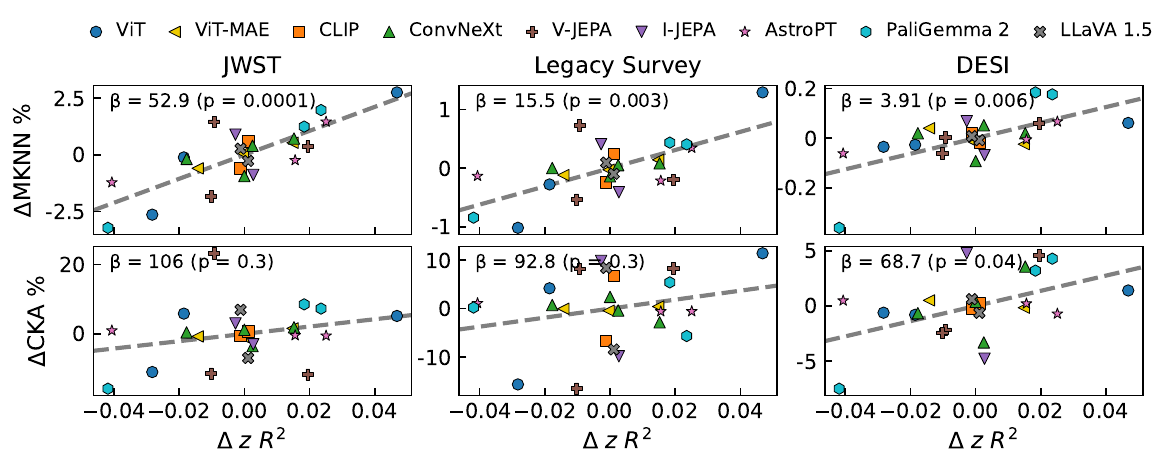}
    \caption{Redshift $(z)$.}
    \end{subfigure}
    \caption{Crossmodal embedding alignment between the stated modality and HSC plotted against probe performance for our tested physical properties.
    ANCOVA $\beta$ and $p$ values are stated for each panel.
    In all tested cases we find a significant correlation between MKNN and physics probe performance.
    We only find a significant correlation between CKA and physics probe performance for DESI Spectra vs HSC.
}
\label{fig_crossmodal_app}
\end{figure}

We list the full results across pairs of data modalities for each model variant in Tab.~\ref{tab_crossmodal_app}. 
Each entry is the MKNN or CKA score as a percentage.
Fig.~\ref{fig_crossmodal_app} plots cross modal MKNN and CKA against linear-probe $R^2$ for each of redshift, stellar mass, and sSFR individually.

\begin{table*}[htbp]
  \caption{
  Crossmodal embedding alignment between a model's embeddings of different astronomical modalities and its embeddings of HSC imaging, measured by MKNN and CKA scores.
  The DESI spectra column uses Specformer embeddings for the DESI side compared against each listed vision model's HSC image embeddings.
  The PRH predicts that both MKNN and CKA scores will increase as models within a family grow in size, since larger models should produce embeddings closer to the Platonic ideal representation.
  \vspace{0.5em}
  }
  \label{tab_crossmodal_app}
  \centering
  \small
  \setlength{\tabcolsep}{6pt}
  \begin{tabular}{l ccc ccc}
    \toprule
    & \multicolumn{3}{c}{MKNN (\%)} & \multicolumn{3}{c}{CKA (\%)} \\
    \cmidrule(lr){2-4} \cmidrule(lr){5-7}
    Model & JWST & Legacy & DESI & JWST & Legacy & DESI \\
    \midrule
    AstroPTv2 Small      & 11.62 & 3.36 & 1.25 & 43.4 & 86.6 & 45.66 \\
    AstroPTv2 Base       & 12.60 & 3.28 & 1.32 & 42.0 & 84.9 & 45.39 \\
    AstroPTv2 Large      & 14.30 & 3.84 & 1.41 & 41.9 & 84.9 & 44.47 \\
    \addlinespace
    CLIP Base            & 12.88 & 1.79 & 1.29 & 30.59 & 59.32 & 33.39 \\
    CLIP Large           & 14.07 & 2.28 & 1.24 & 31.89 & 72.73 & 33.85 \\
    \addlinespace
    ConvNeXtv2 Nano      & 11.33 & 1.13 & 1.01 & 32.52 & 54.06 & 32.29 \\
    ConvNeXtv2 Tiny      & 11.91 & 1.18 & 1.05 & 28.57 & 52.93 & 29.64 \\
    ConvNeXtv2 Base      & 10.57 & 0.99 & 0.87 & 33.28 & 55.73 & 33.32 \\
    ConvNeXtv2 Large     & 12.23 & 1.21 & 1.01 & 34.01 & 50.49 & 36.50 \\
    \addlinespace
    DINOv3 vits16        & 14.55 & 2.42 & 0.98 & 52.18 & 70.23 & 34.80 \\
    DINOv3 vits16plus    & 13.09 & 2.09 & 0.97 & 48.53 & 68.06 & 35.39 \\
    DINOv3 vitb16        & 14.03 & 2.48 & 0.92 & 49.44 & 70.29 & 33.25 \\
    DINOv3 vitl16        & 11.80 & 1.70 & 0.79 & 45.31 & 66.07 & 30.14 \\
    DINOv3 vith16plus    & 10.35 & 1.16 & 0.67 & 31.58 & 43.12 & 22.21 \\
    DINOv3 vit7b16       & 12.62 & 1.89 & 0.82 & 40.14 & 41.65 & 30.13 \\
    \addlinespace
    IJEPA Huge           &  9.85 & 1.27 & 0.56 & 15.22 & 31.64 & 15.86 \\
    IJEPA Giant          & 11.63 & 2.09 & 0.73 & 21.25 & 51.31 & 25.40 \\
    \addlinespace
    LLaVA-1.5 7B         & 11.16 &   1.67 & 1.04 & 31.10 &    56.45 & 36.88 \\
    LLaVA-1.5 13B        & 11.70 &   1.86 & 1.06 & 45.03 &    73.30 & 38.11 \\
    \addlinespace
    PaliGemma 2 3B       & 11.50 & 1.84 & 1.18 & 43.20 & 58.58 & 32.89 \\
    PaliGemma 2 10B      & 12.23 & 1.81 & 1.17 & 41.98 & 47.51 & 33.97 \\
    PaliGemma 2 28B      &  7.03 &  0.56 & 0.50 & 31.2 &  53.39 & 22.21 \\
    \addlinespace
    ViT-MAE Base         & 10.09 & 1.89 & 1.01 & 59.34 & 94.89 & 29.25 \\
    ViT-MAE Large        & 10.75 & 1.98 & 0.94 & 59.23 & 94.49 & 28.39 \\
    ViT-MAE Huge         & 11.22 & 2.15 & 0.93 & 61.66 & 95.35 & 28.59 \\
    \addlinespace
    ViT Base             & 10.48 & 1.20 & 1.02 & 30.13 & 47.35 & 35.17 \\
    ViT Large            & 13.01 & 1.94 & 1.03 & 47.00 & 67.19 & 34.98 \\
    ViT Huge             & 15.88 & 3.51 & 1.14 & 46.32 & 74.42 & 37.18 \\
    \addlinespace
    V-JEPA2 ViT-L        & 14.27 & 3.04 & 0.82 & 59.63 & 77.48 & 23.98 \\
    V-JEPA2 ViT-H        & 10.98 & 1.77 & 0.74 & 24.86 & 52.76 & 23.71 \\
    V-JEPA2 ViT-G        & 13.19 & 2.11 & 0.89 & 24.57 & 77.59 & 30.73 \\
    \bottomrule
  \end{tabular}
\end{table*}

\clearpage

\section{Extended cross-architectural results} \label{app_crossarch}

\begin{figure}[htbp]
    \begin{subfigure}{0.49\textwidth}
        \centering
        \includegraphics[width=\textwidth]{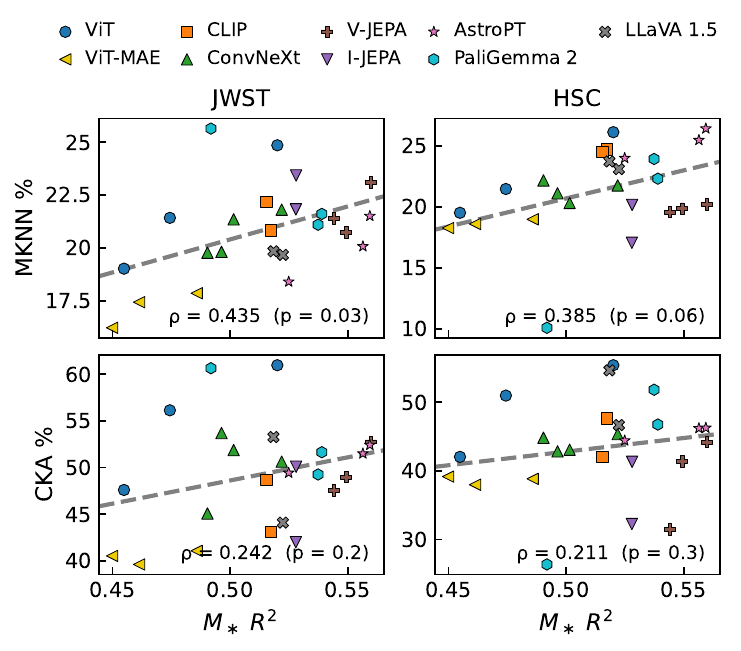}
    \caption{Mass $(M_\star)$.}
    \end{subfigure}
    \begin{subfigure}{0.49\textwidth}
        \centering
        \includegraphics[width=\textwidth]{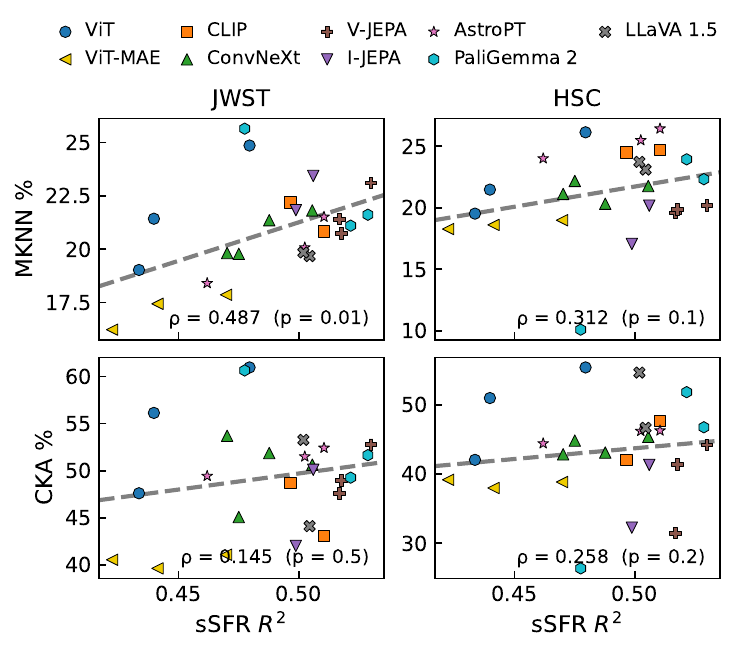}
    \caption{Specific Star Formation Rate (sSFR).}
    \end{subfigure}
    \begin{subfigure}{1\textwidth}
        \centering
        \includegraphics[width=0.49\textwidth]{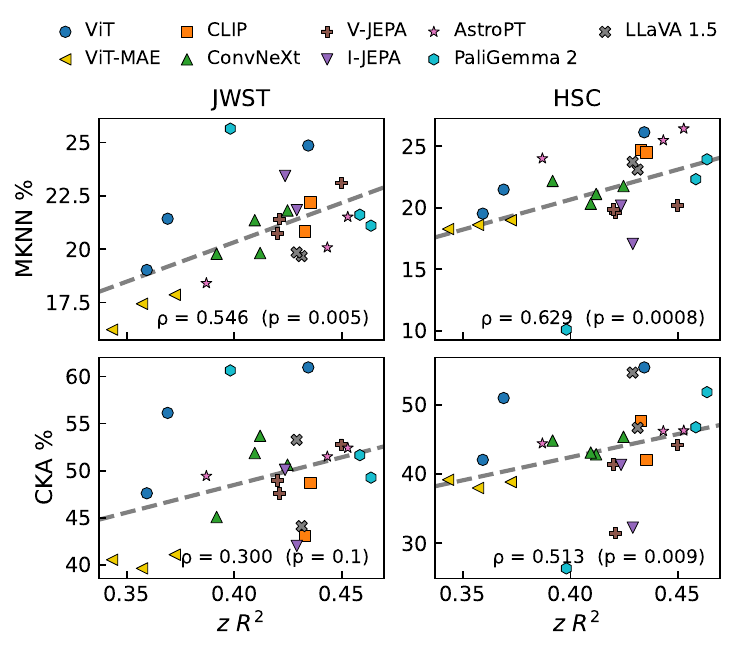}
    \caption{Redshift $(z)$.}
    \end{subfigure}
    \caption{Cross-architectural embedding alignment between the stated modality and HSC plotted against probe performance for our tested physical properties.
    Spearman's $\rho$ and $p$ values are stated for each panel.
    In all tested JWST cases we find a significant correlation between MKNN and physics probe performance.
    For HSC we find significant or borderline correlations between MKNN and physics probe performance.
    We only find a significant correlation between CKA and physics probe performance for HSC redshift.
}
\label{fig_crossarch_app}
\end{figure}

This appendix expands the cross-architectural analysis. 
Figure~\ref{fig_crossarch_app} plots each model's mean MKNN and CKA against the rest of the basket against its linear-probe $R^2$ for redshift, stellar mass, and sSFR individually, on both JWST and HSC imagery.

\subsection{Intervening on the shared physics subspace}
\label{app_causal}

The analyses in \S\ref{subsec_mknncka} show that local embedding alignment tracks physics-probe performance across model families and astronomical modalities. These relationships are nevertheless observational: a third property of the
representations could in principle improve both physical recovery and embedding similarity. We therefore perform a direct intervention on the embedding spaces to ask whether the physical content identified by our probes itself contributes
to representational alignment.

We test 21 pairs drawn from the ViT and ConvNeXtv2 families on HSC imagery. We choose these two families as both are supervised ImageNet-pretrained vision models, reducing variation in the pretraining regime. For each model we fit the
three linear probes described in \S\ref{subsec_probes}, corresponding to redshift ($z$), stellar mass ($M_\star$), and sSFR. Because these quantities are not independent (Fig.~\ref{fig_avg_cos_sim_matrix}), we treat their probe directions
jointly and remove the subspace spanned by all three from each model's embeddings. We then recompute MKNN and CKA between each pair after the intervention.

Our relevant control is not about removing information from the embedding: any low-rank projection may perturb its geometry. We therefore repeat the experiment after removing a random subspace of the same rank. If the alignment
observed in \S\ref{subsec_mknncka} is specifically supported by shared physical content, removing the probe-defined physics subspace should reduce similarity more than removing an equivalently sized random subspace.

\begin{table*}[htbp]
    \centering
    \small
    \caption{
    Effect of removing the probe-defined physics subspace from model embeddings. Values give the mean [min, max] percentage decrease in alignment over 21 tested model pairs. Negative values indicate that alignment increased after the intervention. The random control removes a subspace of equal rank.
    }
    \label{tab:physics_subspace_intervention}
    \setlength{\tabcolsep}{8pt}
    \begin{tabular}{lcc}
        \toprule
        & Physics subspace & Random subspace \\
        \midrule
        CKA (global) & $21\%~[9,\,28]$ & $0\%~[-0,\,0]$ \\
        MKNN (local) & $7\%~[2,\,12]$ & $-7\%~[-12,\,-2]$ \\
        \bottomrule
    \end{tabular}
\end{table*}

Removing the physics subspace reduces MKNN for every tested pair, with a mean decrease of 7\% and a range of 2--12\%. Removing a random subspace of the same rank has the opposite effect: MKNN increases by 7\% on average. We find an even
larger change in absolute CKA, which decreases by 21\% after removing the physics subspace while remaining effectively unchanged under the random control.

This intervention answers a different question from the scaling experiments in \S\ref{subsec_mknncka}. The latter ask whether representational similarity \emph{tracks} physics performance as models improve, for which we consistently
find a relationship for MKNN but not CKA. Here we instead ask whether the probe-defined physical content makes a direct contribution to the similarity between two fixed embedding spaces. The decrease following targeted removal,
together with the equal-rank random control, shows that the shared physics subspace is not merely correlated with the observed alignment: intervening on that subspace directly changes it. This provides causal evidence that the
astrophysical content recovered by our probes contributes to the representational convergence measured in the main experiments.

\end{document}